\documentclass{aa}  

\usepackage{graphicx}
\usepackage{array} 
\usepackage{subfig}
\usepackage{placeins}
\usepackage{float}
\usepackage{afterpage}
\usepackage{multirow}
\usepackage{rotating}

\usepackage{txfonts}
\usepackage{tikz,xcolor,hyperref}
\definecolor{lime}{HTML}{A6CE39}
\DeclareRobustCommand{\orcidicon}{%
	\begin{tikzpicture}
	\draw[lime, fill=lime] (0,0) 
	circle [radius=0.16] 
	node[white] {{\fontfamily{qag}\selectfont \tiny ID}};
	\draw[white, fill=white] (-0.0625,0.095) 
	circle [radius=0.007];
	\end{tikzpicture}
	\hspace{-2mm}
}

\usepackage{txfonts}
\foreach \x in {A, ..., Z}{%
	\expandafter\xdef\csname orcid\x\endcsname{\noexpand\href{https://orcid.org/\csname orcidauthor\x\endcsname}{\noexpand\orcidicon}}
}

\newcommand{\bene}{\textcolor{blue}}

\begin{document}

   \title{Measuring the IGM correlation length at 5<z<6.1: a fast change at the end of Reionization}

  \author{Benedetta~Spina\inst{1}\fnmsep\thanks{b.spina@thphys.uni-heidelberg.de}\orcidA{},
          Sarah~E.~I.~Bosman\inst{1,2}\orcidB{},  
          Barun~Maity\inst{2}\orcidD{} \and Frederick~B.~Davies\inst{2}\orcidC{} 
          }
   \authorrunning{Spina et al.}
   \institute{Institute for Theoretical Physics, Heidelberg University, Philosophenweg 12, D–69120, Heidelberg, Germany
         \and
             Max-Planck-Institut f\"{u}r Astronomie, K\"{o}nigstuhl 17, 69117 Heidelberg, Germany
             }
             
   \date{}

  \abstract{
  The Lyman-$\alpha$ forest of high redshift quasars is a powerful probe of the late stages of the Epoch of Reionization (EoR), in particular thanks to the presence of Gunn-Peterson troughs. These troughs exhibit a broad range of lengths, with some extending up to $\sim 100$ cMpc, suggesting large-scale coherent structures in the intergalactic medium (IGM).}{We aim to gain more insight into the presence, extent, and magnitude of correlations in the Lyman-$\alpha$ forest during the end of reionization, at $5<z<6.1$. In particular, we want to quantify the scales over which correlations are significant in order to help inform the required size of cosmological simulations aiming to capture the evolution of the EoR.}{We utilize the extended XQR-30 dataset over the redshift range $5 < z < 6.1$ ($\Delta z = 0.05$) to explore large-scale correlations. After accounting for the relevant systematics, the flux correlation matrix proves to be a powerful tool for probing large-scale correlations across redshifts. We perform an MCMC analysis to quantify the extent and strength of the correlation, making use of several functional forms. We moreover employ new large-volume ($1.5^3\,\mathrm{Gpc}^3$) light-cones of Lyman-$\alpha$ transmission implementing different reionization scenarios to interpret the observed signal, including a fiducial box employing SCRIPT.}{We detect strong correlations at redshifts $z > 5.3$, extending at least tens of Mpc and strongly increasing with redshift. Our results suggest a redshift-dependent correlation length, from $L \leq 26.53\, (68.47)\,\mathrm{Mpc}$ at 1-$\sigma$ (2-$\sigma$) limit at redshift $z = 5.0$ to $L = 252.72^{+272.61}_{-41.61}\,\mathrm{Mpc}$ at redshift $z = 6.1$. On the contrary, our simulations all demonstrate characteristic correlation scales $< 60$ Mpc with a very slow redshift evolution, in strong tension with our observations.}{The presence and redshift-dependence of correlations in the Lyman-$\alpha$ forest on $>200$ Mpc scales at $z=6$ indicates that cosmological simulations should be larger than this scale to adequately sample the Lyman-$\alpha$ forest. Despite implementing a fluctuating UVB and numerous neutral islands at $z<6$, and matching well the sightline scatter in the Lyman-$\alpha$ forest, our fiducial SCRIPT-based simulation fails to reproduce the large-scale correlations. It may be that those ingredients are necessary, but not sufficient, to understanding the unfolding of the EoR.}
  

   \maketitle
%
%-------------------------------------------------------------------

\section{Introduction}

The Epoch of Reionization (EoR) marks the last major global phase transition in the history of the Universe. During this period, the Intergalactic Medium (IGM) gradually became ionized while structures formed, making the study of the EoR crucial to understand IGM physics and galaxy formation in the early universe. 

The midpoint of the EoR is set at redshift $z \sim 7-8$ by observations of the Cosmic Microwave Background \citep{Planck18}, the decline of Lyman-$\alpha$ emission from galaxies at $z > 6$ \citep{Mason18,Jung20,Wold22}, the analyses of the Lyman-$\alpha$ damping wing in quasars at $z > 7$ \citep{Wang20,Yang20b,Greig22} and the IGM thermal state measurements at $z > 5$ \citep{Gaikwad20, Gaikwad21}. 
It is in this context that the analysis of high quality spectra obtained from high redshift quasars (QSOs) emerges as an essential tool to probe the timeline of the EoR and its end in particular, starting with the measurement of the fraction of dark pixels in the Lyman series forests \citep{McGreer15}. 

Although substantial sightline-to-sightline variations in Lyman-$\alpha$ optical depth can arise from a decline in the ionizing background, even without a significantly neutral intergalactic medium \citep{Lidz06,BoltonHaehnelt07,Davies16,D'Aloisio18}, the observed scatter exceeds expectations from cosmic density fluctuations alone \citep{Becker15,Bosman18,Yang20,Zhu21,Bosman22}, suggesting a late and patchy reionization scenario and placing the end of the EoR at $z \simeq 5.3$ \citep{Kulkarni19,Nasir20,Keating20}. This picture is also supported by the rapid evolution of the mean free path of ionizing photons at $5 < z < 6$ \citep{Becker21, Bosman21SA, Zhu23}, and the underdensities around long gaps traced by Lyman-$\alpha$ emitting galaxies \citep{Becker18, Christenson23}.

A key constraint provided by the Lyman-$\alpha$ forest arises from the Gunn-Peterson (GP) effect — the absorption of rest-frame (wavelength of $1215.67$ \AA) quasar photons by neutral hydrogen \citep{GP}, with an HI fraction $x_\mathrm{HI} \gtrsim 10^{-4}$ along the line of sight to background quasars —leading to the so-called Gunn-Peterson troughs. In particular, the detection of an extended ($\sim 110 \,h^{-1}\mathrm{Mpc}$) trough in ULAS J0148+0600 \citep{Becker15} challenged the previously-held scenario that reionization ended around $z \sim 6$ and highlighted the potential of GP troughs to probe the final stages of the EoR. 
Since Gunn-Peterson troughs appear even for very low $x_{\mathrm{HI}}$ \citep{Mesinger10}, their presence alone does not distinguish between residual neutral gas and large, significantly neutral islands. However, the detection of damping wings around GP troughs in the Lyman-$\alpha$ forest of $z < 6$ quasars \citep{Spina+24, Zhu24} suggests that substantial neutral hydrogen persisted late.

Gunn-Peterson troughs have been detected with sizes extending for tens of comoving Mpc, indicating extended coherent neutral regions in the IGM, and implying the existence of correlations on scales approaching 100 comoving Mpc. If this is the case, what is the reason for such large coherent regions? What can we infer on the physics driving the EoR from these observation? And from a more practical point-of-view, how reliable are simulations with box sizes smaller than this scale in accurately representing the evolution of the EoR? How large must such simulation boxes be to fully capture these correlations? The purpose of this work is to address these questions and constrain the characteristic correlation scale.

The idea that the reionization process affects and has structure on all scales was shown in \citet{Furlanetto16} by analyzing the EoR through percolation theory, but without quantifying strength or impact, i.e.~whether it would be detectable. 
Whether $\sim100$ comoving Mpc simulations are sufficient for the convergence of key reionization features—such as the size and distribution of ionized bubbles—is extensively discussed in \citet{Gnedin22}, suggesting the need for simulations larger than 250 comoving Mpc. The issue of convergence in simulation boxes — and, more broadly, the correlation of scales — has been poorly explored from a theoretical perspective, and it remains almost unaddressed in observations. The first attempt to do so appears in \citet{Fan+06b, Fan06}, where the evolution of the IGM optical depth is investigated and large fluctuations in the UV background are reported, implying that transmission is correlated over large scales. By accounting for the redshift evolution across different sightline bins in a sample of 19 quasars, they reported a correlation scale of tens of comoving Mpc.
From a theoretical perspective, similar conclusions were reached by \citet{Battaglia13}, by measuring the power spectrum of the ionization field in large-scale EoR simulations, and by \citet{Qin21} cross-correlating the overdensity field with the Lyman-$\alpha$ transmission flux. Previous attempts \citep[e.g.,][]{Croft08} in quantifying the correlation of scales introduced by the reionization process have been limited by the small volume accessible from high-resolution simulations and have been mainly focused on estimating the size and distribution of ionized bubbles \citep{Furlanetto06,Furlanetto09}. Studies on the Lyman-$\alpha$ forest flux autocorrelation function at $z \geq 5$ \citep{Wolfson23} have pointed out significant correlations at small scales (of the order of a few Mpc), without addressing larger scales. \citet{Zhu21}, on the other hand, focus on the occurrence rates of GP troughs alone, which are rare especially at $z<5.5$. In contrast, we will address the correlation scales on the transmission directly.

We aim to provide a systematic study of the optical depth correlation towards the end of the EoR using high-resolution QSOs spectra at high redshift from the E-XQR-30 quasar sample \citep[][]{D'Odorico23}, obtained with the X-Shooter instrument \citep[][installed
at the ESO-VLT telescope]{Vernet11} and ESI \citep[][installed
at the Keck II telescope]{Sheinis02} instruments. The sample is fully described in \citet{Bosman22}.

The paper is organized as follows. In Section~\ref{sec:sample}, we describe our observational sample, detailing the continuum-reconstruction procedure (Section~\ref{sec:continuum}) and addressing key systematics in the data. Our methodology is outlined in Section~\ref{sec:methods} and it includes the data preparation for fitting (Section~\ref{sec:correlation_matrix}) and the functional forms employed (Section~\ref{sec:functional_forms}). We present and analyze our results in Section~\ref{sec:results}. In Section~\ref{sec:simulations} we introduce and investigate the simulations used to interpret our findings, followed by the discussion and conclusions in Section~\ref{sec:discussion}. The cleaned and binned QSO spectra are shown in Appendix~\ref{sec:QSO_list}, while details on the Monte Carlo Markov Chains posteriors are shown in Appendix~\ref{sec:MCMC}, and further comparisons between data, models and simulations are displayed in Appendix~\ref{sec:bigbox_fit}. Throughout the paper we use a Planck flat $\Lambda$CDM cosmology \citep{Planck18}, and distances are given in comoving units.

\begin{figure*}
    \centering
    \centerline{\includegraphics[width=\textwidth]{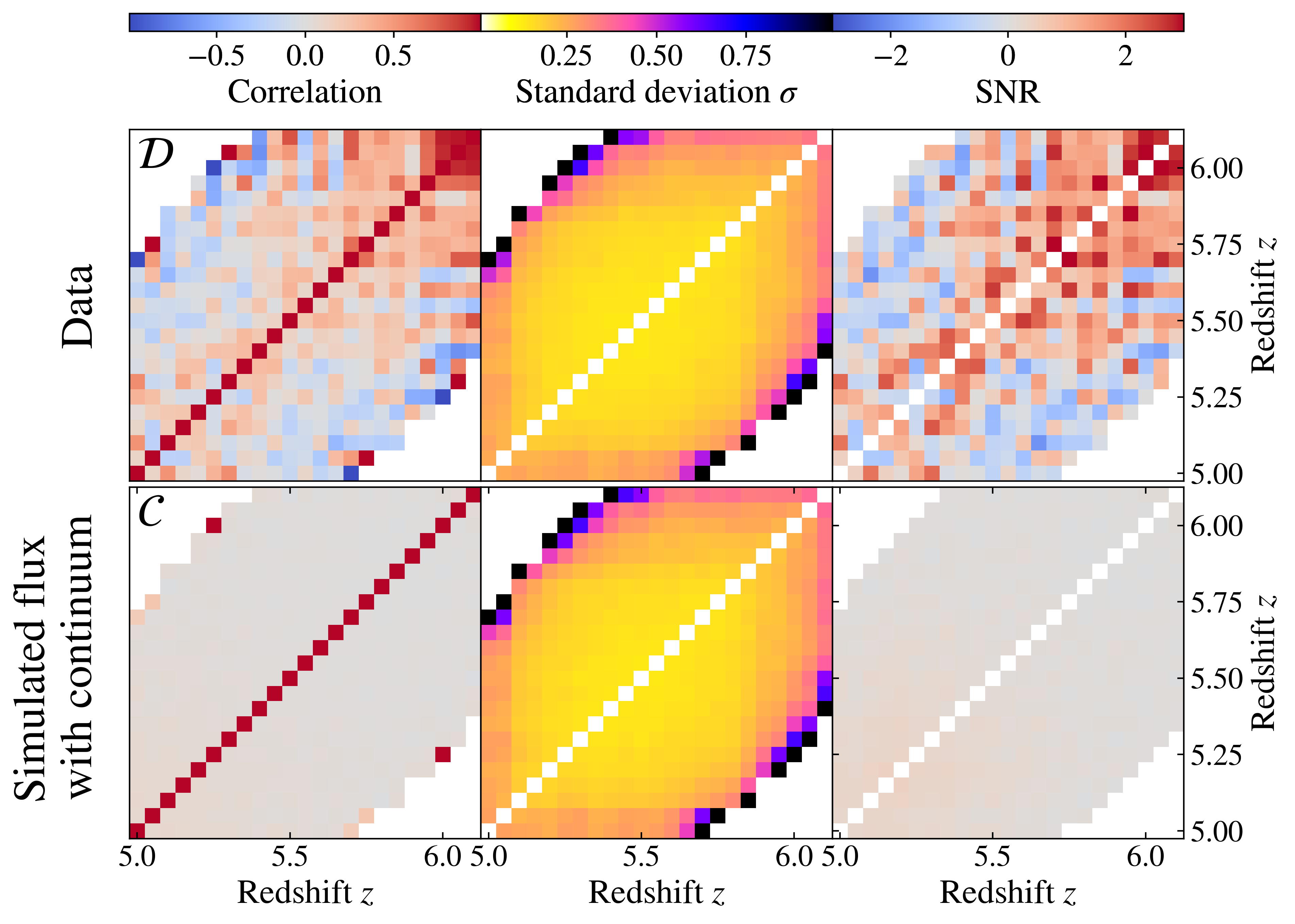}}
    \caption{Correlations in the Lyman-$\alpha$ forest flux across redshift. Top row: correlation matrix for observed QSO sightlines, showing a positive correlation (in red) in the off-diagonal terms near the diagonal at all redshifts, and an increasing correlation length at higher redshifts. Bottom row: mock sightlines with continuum-reconstruction uncertainties applied, leading to a slight increase in correlation at lower redshifts, while failing to account for the observed correlation at higher redshifts. First column: correlation matrices. Second column: standard deviation of the correlation matrices. Third column: statistical significance of the correlation, estimated as the ratio between the correlation matrix and its standard deviation. The results suggest that while continuum-reconstruction uncertainties contribute to correlations at low redshifts, they cannot account for the correlations observed at higher redshifts.}
    \label{fig:fig2}
\end{figure*}

\section{Sample} \label{sec:sample}
We use 67 QSO spectra at redshift $z > 5.5$, with signal-to-noise ratios $\mathrm{SNR} \geq 10$ per $\leq 15\,\mathrm{km\,s}^{-1} $ spectral pixel. We combine observations from the X-Shooter and ESI instruments in order to increase our sample size and improve the statistical analysis. 

For each sightline, the flux is binned into redshift bins of $\Delta z = 0.05$  in the redshift range $4.975 \leq z \leq 6.125$ (the redshift bins are therefore centered at $z_\mathrm{bin} = 5.00, 5.05, \dots, 6.10$), covering the last stage of the EoR and entering the post-EoR regime. We include non-detections in our sample as well (the flux is hence allowed to be negative), since both the observational uncertainties and the uncertainties related to the continuum-reconstruction are available and well-behaved.
The redshift bins for which the Lyman-$\alpha$ forest is available are peculiar to each QSO, and as expected, not all the QSOs contribute to all the redshift bins. In particular, the redshift bins closer to the edges our redshift range are under-sampled with respect to the central bins. It also implies that not all of the redshift combinations (i.e.~redshift 5 with redshift 6) provides constraints with the same robustness. We apply the same masking on the blue side of the Lyman-$\alpha$ emission line—to account for proximity zone contamination—as in \citet{Bosman22}, excluding  rest-frame wavelengths $\lambda > 1185$ \AA. Damped Lyman-$\alpha$ systems (DLAs) identified along quasar sightlines \citep{DaviesR23} are masked. A more detailed description of the sample and the data-cleaning procedures (e.g.~masking of residuals from sky emission lines) can be found in \citet{Bosman22}.

In addition, we remove those redshifts which might be contaminated by the OVI doublet emission line ($\lambda\lambda 1032, 1038$\AA), i.e.~closer than $\Delta v = 5,000\, \mathrm{km\,s^{-1}}$ from the OVI redshifted peak. This conservative mask reduces the redshift bins available for the analysis but robustly prevents any contamination that might introduce artificial correlations, since OVI is a more difficult emission line to reconstruct in quasar spactra than the other line in the wavelength range we use \citep{Bosman21}. The effects of the masking are shown in Appendix~\ref{sec:QSO_list}, along with the wavelengths used for the analysis.

The full sample is presented in Figure~A.1. For each quasar, the binned flux is presented in black as a function of redshift (the gray shadow region represent the continuum-reconstruction error) and the data selected by accounting for the OVI doublet emission line (semi-dashed light blue vertical line) in red. The Lyman-$\alpha$ and Lyman-$\beta$ lines are also displayed as the dashed orange and dotted blue lines, respectively.

\subsection{Continuum reconstruction} \label{sec:continuum}
The Lyman-$\alpha$ forest of each quasar is normalized by reconstructing the continuum flux using a near-linear log-PCA method \citep{Davies18, Bosman21, Bosman22}. This approach has been shown to predict the underlying continuum within an $\sim 8\%$ uncertainty for each quasar, though this uncertainty is highly coherent (i.e.~the entire continuum can be systematically under- or overestimated).
To account for this uncertainty, we simulate the effect of the continuum-reconstruction using mock observations of the Lyman-$\alpha$ forest from \citet{Davies24}. The mock observations consists of 1000 independent realizations of the QSO sample presented in \citet{Bosman22}. Furthermore, they are tailored to match the observed sample e.g. in terms of the number of quasars contributing to each redshift bin, and are obtained by recalculating the original redshift binning of $\Delta z = 0.1$  to the redshift binning used in this work, $\Delta z = 0.05$. Specifically, for each redshift and each quasar in the mock sample, the continuum-normalized flux is perturbed by a factor of $1+\mathcal{G}(\sigma_{z, \,\mathrm{QSO}}^\mathrm{cont})$, where $\mathcal{G}(\sigma_{z, \,\mathrm{QSO}}^\mathrm{cont})$ is a random number drawn from a Gaussian distribution with a standard deviation specific to each quasar at each redshift \citep[measured in][]{Bosman22}, representing the uncertainty on continuum-reconstruction. 
This approach carefully and accurately models the continuum-reconstruction process by systematically varying the flux across redshift bins, enabling the generation of realistic mock spectra that reflect observational properties. It is worth noting that by applying this procedure we are conservatively assuming a complete error correlation across all scales, slightly overestimating by $\sim10\%$, on purpose, the error introduced by the continuum reconstruction on large scales.

\section{Methods} \label{sec:methods}

We aim to investigate and quantify the strength and extent of Lyman-$\alpha$ forest correlations on different scales. We achieve this by measuring the correlation between combinations of redshift bins and evaluating its statistical significance, accounting for systematics such as continuum reconstruction and observational uncertainties.

\begin{figure*} 
    \centering
    \subfloat[Model $\mathcal{M}1$: constant\label{fig:fig_constant}]{\includegraphics[width=0.48\textwidth]{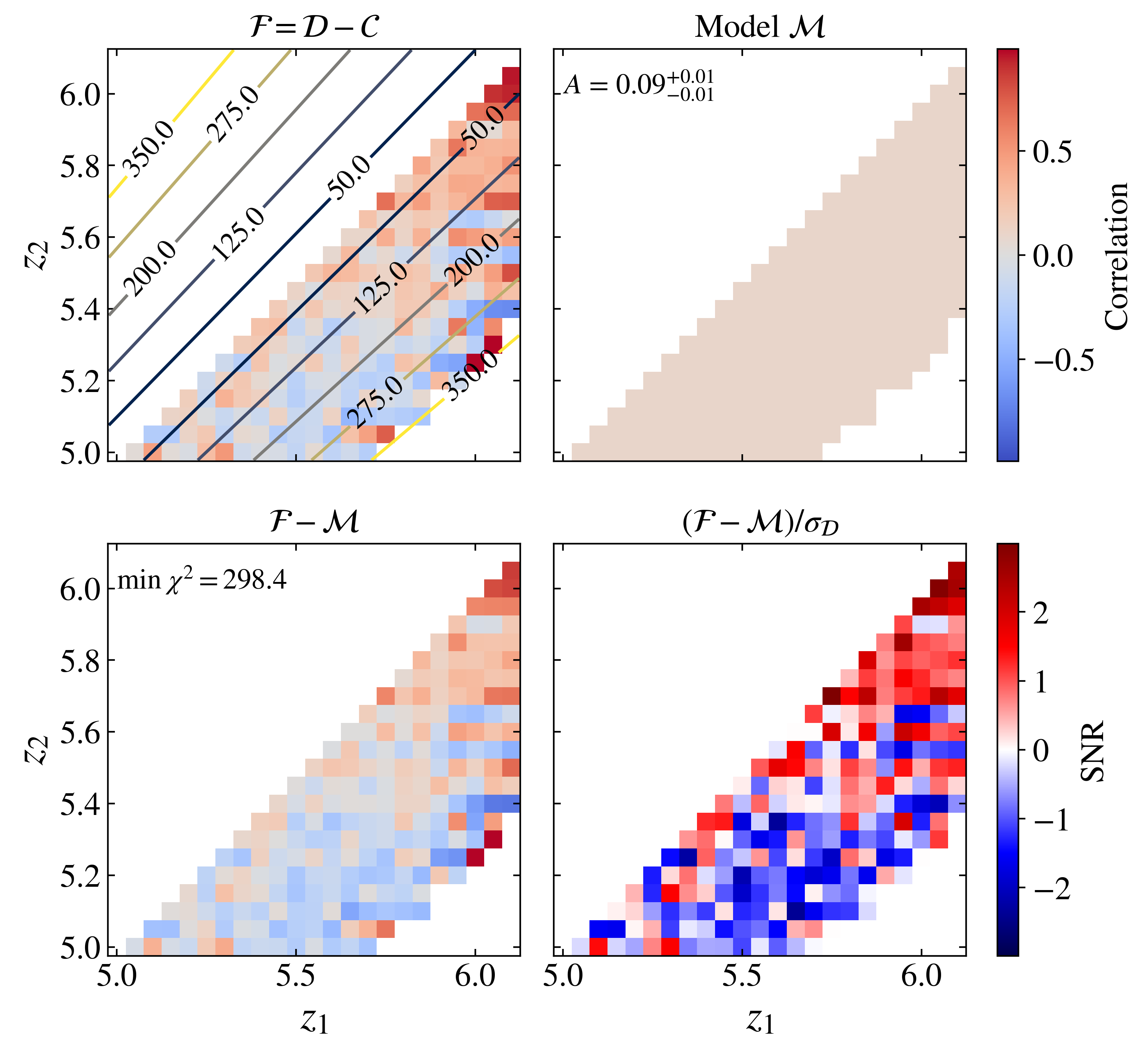}}\hfill
    \subfloat[Model $\mathcal{M}2$: gaussian, constant amplitude and width\label{fig:fig_gaussian}]{\includegraphics[width=0.48\textwidth]{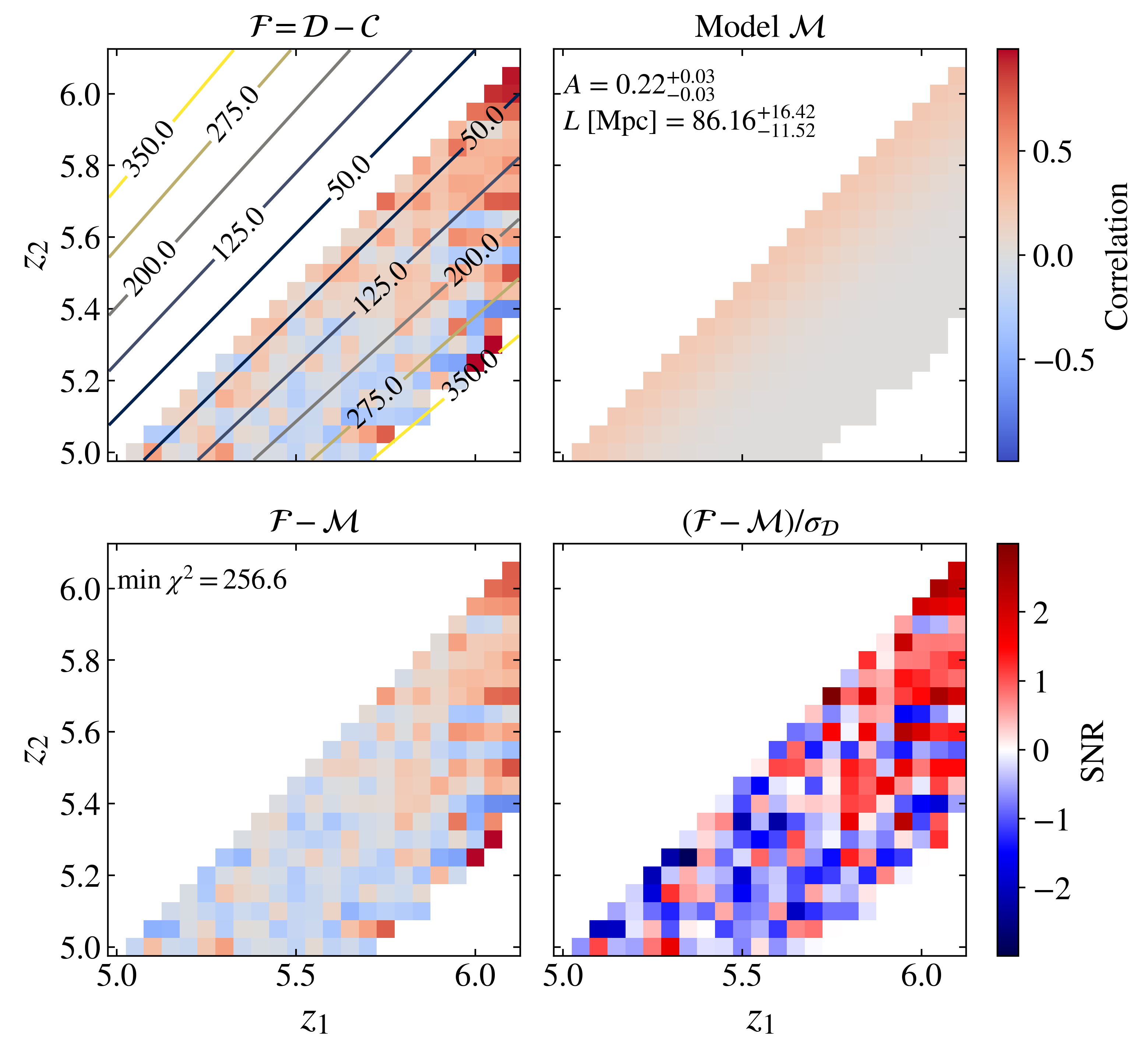}}\\
    \subfloat[Model $\mathcal{M}3$: gaussian, amplitude and width redshift-dependents\label{fig:fig_general}]{\includegraphics[width=0.48\textwidth]{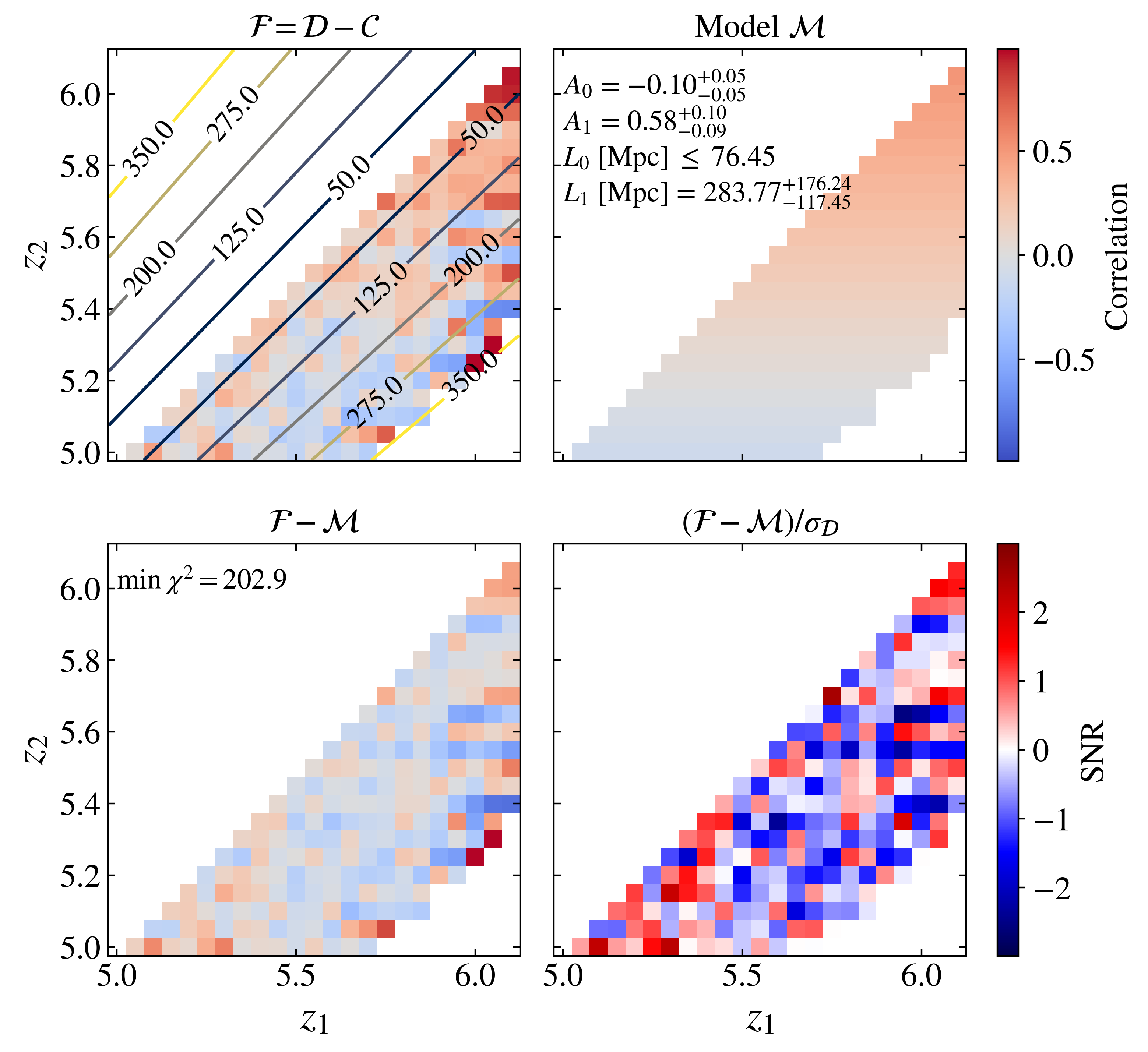}}\hfill
    \subfloat[BIBORTON SCRIPT\label{fig:fit_S3}]{\includegraphics[width=0.48\textwidth]{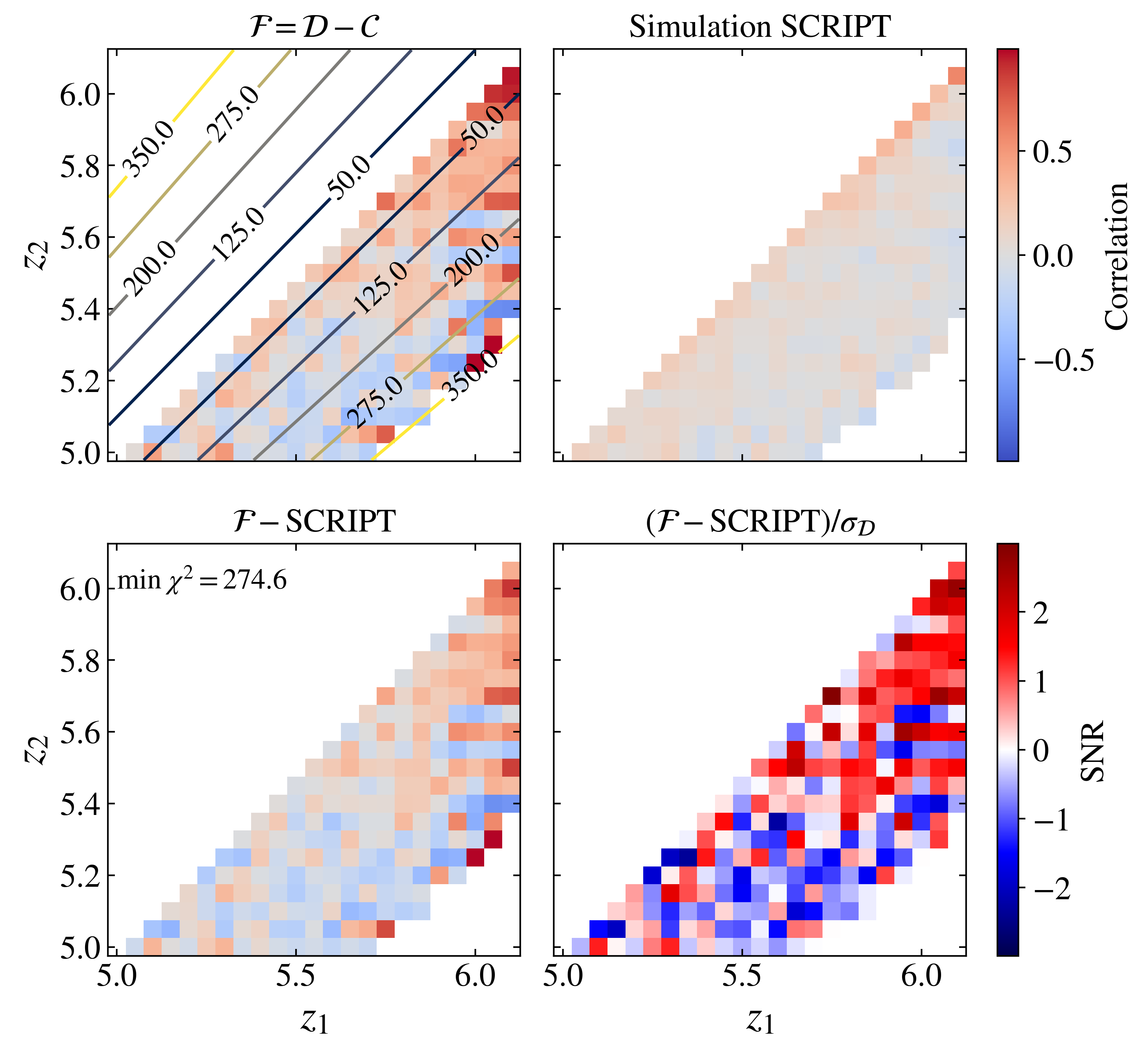}}
    \caption{Results of the fitting for different models model. For each model, we show on the \textit{top-left panel} the correlation matrix used for the fitting, obtain by removing the effects from the continuum-reconstruction from the data; isocontours corresponding to the comoving distances in units of Mpc at each redshift pair are also shown. The \textit{top-right panel} displays results from the best fitting of the model here considered, $\mathcal{M}$. The residuals between the data and the best fit of the model, $\mathcal{F}-\mathcal{M}$ are shown in the \textit{bottom-left panel}. Finally, the \textit{bottom-right panel} shows the signal-to-noise ratio of the data. Analogue panels for the simulations BIBORTON $\mathrm{const}\lambda$ and $\mathrm{evolv}\lambda$ are displayed in Figure~\ref{fig:fit_S_appendix}.}
    \label{fig:fit_models_to_data}
\end{figure*}

\subsection{Correlation matrix} \label{sec:correlation_matrix}

As a first step, we compute the correlation matrix $\mathcal{D}$ of the sightlines described in Sec.~\ref{sec:sample}, shown in the top-left panel of Figure~\ref{fig:fig2}. Let $S_i$ denote the flux transmission field along the $i$-th QSO sightline, discretized into $M$ redshift bins, with $i = 1, \dots, N$ and $N = 67$ the total number of sightlines considered. We compute the sample mean $\bar{S}$ for each redshift bin, the elements of the covariance matrix $\mathsf{C}_{jk}$ (where $j,k = 1, \dots, M$ denote the redshift bins), and the elements of the correlation matrix $\mathcal{D}_{jk}$ of the sightline ensemble as follows:

\begin{align}
\bar{S} &= \frac{1}{N} \sum_{i=1}^N S_i, \\
\mathsf{C}_{jk} &= \frac{1}{N - 1} \sum_{i=1}^N \left( S_i^{(j)} - \bar{S}^{(j)} \right) \left( S_i^{(k)} - \bar{S}^{(k)} \right), \\
\mathcal{D}_{jk} &= \frac{\mathsf{C}_{jk}}{\sqrt{\mathsf{C}_{jj} \, \mathsf{C}_{kk}}}.
\end{align}

A positive correlation is consistently observed at all redshifts in the off-diagonal terms near the diagonal and even in more distant terms at higher redshifts, suggesting that correlations extend not only between neighboring bins but also across redshift separations of up to $\delta z\sim 0.5$. An enhanced correlation at high redshift would imply a scale-coherent behavior of the IGM during the late stages of the EoR, indicating a more interconnected structure than in the post-reionization era.

However, whether this correlation is statistically significant remains to be determined, as it could arise from systematic effects, such as those introduced by the continuum-reconstruction (here conservatively assumed to be fully correlated). 
To assess the strength of the signal, we shuffle the flux in the sightlines by randomly permuting the flux values across sightlines within each redshift bin $N = 1000$ times and compute the standard deviation of the correlation matrices produced in each shuffling, effectively computing the standard deviation of  $\mathcal{D}$, $\sigma_\mathcal{D}$, presented in the top-center panel of Figure~\ref{fig:fig2}. As previously noted, the number of QSOs contributing to each redshift combination decreases toward the edges of the considered redshift range, resulting in a higher standard deviation in these regimes.
A first, approximate yet informative measure of the statistical significance of the correlation is given by the ratio between the correlation matrix and its standard deviation, as shown in the top-right panel of Figure~\ref{fig:fig2}. A signal-to-noise ratio SNR $\geq 2$ correlation is visible in the off-diagonal terms near the diagonal across all redshifts, with increasing correlation length at higher redshifts. Meanwhile, the apparent correlation between very low redshifts (e.g., $z = 5$) and very high ones (e.g., $z=5.7$), due to the size of our sample and observed in the top-left panel, appears not to be statistically significant.

\begin{table*}
\centering
\renewcommand{\arraystretch}{1.8}
\begin{tabular}{c p{3.2cm} p{6.7cm} p{2cm} p{2.8cm} }
\hline\hline\hline    

\multirow{5}{*}{\rotatebox{90}{\large{Observations} \hspace{3.2cm}}} & Model & Functional form & Best fit & $\min\chi^2/N_\mathrm{dof} = \chi^2_{N_\mathrm{dof}}$ \\
\hline\hline

& $\mathcal{M}1$: constant & $\mathcal{M}1(z_1,z_2) = A$ & $A = 0.09^{+0.01}_{-0.01}$ & $298.4/220 = 1.35$ \\ 
\cline{2-5}

& $\mathcal{M}2$: Gaussian, constant amplitude and width & 
{\vspace{-1cm}
\begin{equation*}
    \mathcal{M}2(z_1,z_2) = A\,\exp\left[-\dfrac{1}{2}\left(\dfrac{\chi(z_1)-\chi(z_2)}{L}\right)^2\right]
\end{equation*}
\vspace{-0.4cm}
}
& 
{\vspace{-1.1cm}
\begin{align*}
    A & = 0.22^{+0.03}_{-0.03}, \\
    L & = 86.16^{+16.42}_{-11.52}\,\mathrm{Mpc}
\end{align*}
\vspace{-0.6cm}
}& $256.6/219 = 1.17$  \\ 
\cline{2-5}

& $\mathcal{M}3$: Gaussian, redshift-dependent amplitude and width & 
{\vspace{-1cm}
\begin{align*}
    \mathcal{M}3(z_1,z_2) & = A(z_1,z_2)\,\exp\left[-\dfrac{1}{2}\left(\dfrac{\chi(z_1)-\chi(z_2)}{L(z_1,z_2)}\right)^2\right], \\
    A(z_1,z_2) & = A_0+A_1\cdot(z_2-5), \\
    L(z_1,z_2) & = L_0+L_1\cdot(z_2-5) 
\end{align*}
\vspace{-0.5cm}
} & 
{\vspace{-1cm}
\begin{align*}
    A_0 & = -0.10^{+0.05}_{-0.05}, \\
    A_1 & = 0.58^{+0.10}_{-0.09}, \\
    L_0 & \leq 76.45\,\mathrm{Mpc}, \\
    L_1 & = 283.77^{+176.24}_{-117.45}\,\mathrm{Mpc}
\end{align*}
\vspace{-0.5cm}
}& $202.9/217 = 0.94$ \\
\cline{2-5}

& $\mathrm{const}\lambda$ & \mbox{-----------------} & \mbox{---------} & $392.3/221 = 1.76$  \\ \cline{2-5}
& $\mathrm{evolv}\lambda$ & \mbox{-----------------} & \mbox{---------} & $278.8/221 = 1.26$  \\ \cline{2-5}
& $\mathrm{SCRIPT}$ & \mbox{-----------------} & \mbox{---------} & $274.6/221 = 1.24$  \\ \hline\hline\hline

\multirow{2}{*}{\rotatebox{90}{\large{BIBORTON} \hspace{0.75cm}}}  & Model & \multicolumn{2}{c}{Best fit} & $\min\chi^2/N_\mathrm{dof} = \chi^2_{N_\mathrm{dof}}$ \\
\hline\hline

 & $\mathcal{M}3$ with $\mathrm{const}\lambda$ & \multicolumn{2}{l}{$ A_0  = 0.78^{+0.04}_{-0.02},\,A_1  = -0.36^{+0.02}_{-0.06},\, L_0  = 37.81^{+0.47}_{-1.56}\,\mathrm{Mpc},\,  L_1 \leq 3.86 \,\mathrm{Mpc}  $} & $265.8/217 = 1.23$\\ 
\cline{2-5}

 & $\mathcal{M}3$ with $\mathrm{evolv}\lambda$ & \multicolumn{2}{l}{$ A_0  = 0.20^{+0.03}_{-0.02},\,A_1  = 0.17^{+0.02}_{-0.05},\, L_0  = 46.14^{+0.72}_{-4.56}\,\mathrm{Mpc},\,  L_1 \leq 10.80\,\mathrm{Mpc}  $} & $315.4/217 = 1.45$ \\ 
\cline{2-5}

 & $\mathcal{M}3$ with $\mathrm{SCRIPT}$ & \multicolumn{2}{l}{$ A_0  = 0.11^{+0.02}_{-0.02},\,A_1  = 0.19^{+0.03}_{-0.05},\, L_0  = 42.72^{+3.11}_{-4.83}\,\mathrm{Mpc},\,  L_1  \leq 8.28\,\mathrm{Mpc}  $} & $294.8/217 = 1.36$ \\ 
\hline\hline

\end{tabular}
\caption{Summary of models and their best-fit parameters.}
\label{tab:models}
\end{table*}

In order to account for the impact of the continuum reconstruction uncertainties, we repeat the same procedure using the mock observations of the Lyman-$\alpha$ forest described in Sec.~\ref{sec:continuum}, shown in the second row of Figure~\ref{fig:fig2}. These mock sightlines contain no intrinsic correlation across scales, apart from the one introduced by the continuum-reconstruction correlated uncertainties, as shown in their correlation matrix $\mathcal{C}$. A mild but visible correlation appears at redshifts $z\sim 5.0-5.5$ with a persistent and correlated SNR $\sim 0.4$ in the bottom-left panel of Figure~\ref{fig:fig2} when introducing the continuum uncertainties. Its statistical significance $\sigma_\mathcal{C}$ is displayed in the bottom-center panel of Figure~\ref{fig:fig2}.

While uncertainties in continuum-reconstruction may account for the observed correlation in the off-diagonal terms at low redshifts, they do not explain the correlation seen at higher redshifts. To assess the extent of this contribution, we repeated the continuum-reconstruction procedure described in Section~\ref{sec:continuum}, by articially increasing the uncertainty on the continuum reconstruction measured in \citet{Bosman22}, denoted as $\sigma_{z, ,\mathrm{QSO}}^\mathrm{cont}$. While being confident on the accuracy of the reconstruction, we allow for the measured uncertainties to be underestimated by  50\% and we test the sensitivity to the continuum-reconstruction uncertainty of our results.
As expected, this led to an enhancement of the correlation at low redshifts (SNR $\sim 0.8$ ), consistent with the trend already visible in the bottom-left panel of Figure~\ref{fig:fig2}. However, this modification had no apparent effect on the correlation at higher redshifts (consistent within 10\%), reinforcing the robustness of our results against systematic errors. 

\subsection{Fitting through functional forms} \label{sec:functional_forms}
The correlation matrix analysis in Section~\ref{sec:correlation_matrix} suggests a strong correlation at high redshift. We now aim to quantify and model this correlation.
As a first step, we compute the difference between the correlation matrices of the data, $\mathcal{D}$, and of the mock sightlines with continuum uncertainty, $\mathcal{C}$. This subtraction isolates the intrinsic correlation by removing the effects introduced by continuum reconstruction. This approach does not account for the observational uncertainties. These, however, are an order of magnitude smaller than continuum uncertainties and mostly uncorrelated across different redshifts.
The resulting correlation matrix, $\mathcal{F} = \mathcal{D} - \mathcal{C}$, provides the basis for testing different models of correlation scales.

We consider 3 functional forms to model the signal we have detected,
\begin{itemize}
    \item[$\mathcal{M}1$] A constant model, preferred if there is no correlation between scales. \\
    
    \item[$\mathcal{M}2$] A gaussian model, with fixed amplitude and width. It will be statically preferred over $\mathcal{M}1$ if there is correlations are present. In this model, the characteristic correlation scale does not depend on redshift (i.e.~scales are correlated at a fixed distance). The functional form depends on the comoving distance between $z_1$ and $z_2$. \\
    
    \item[$\mathcal{M}3$] A gaussian model, as $\mathcal{M}2$, but with redshift-dependent amplitude and width; in this case the correlation length is not fixed but depends on the redshift considered. It will be preferred over $\mathcal{M}2$ if there is statistically significant evidence for a redshift-dependence of the characteristic correlation scale. 
\end{itemize}

We use Monte Carlo Markov Chain (MCMC) methods to explore the parameter space and fit different models to the correlation matrix $\mathcal{F}$. We perform the MCMC sampling using the \texttt{emcee} package \citep{Foreman-Mackey13}. We use the reciprocal of each term of the standard deviation matrix $\sigma_\mathcal{D}$ to compute the diagonal terms of the inverted covariance matrix, required for the likelihood evaluation. This approach ensures proper weighting of the data uncertainties while fitting the correlation model. 

In the following, we present the functional forms and the results of the fitting in Table~\ref{tab:models} and Figures~\ref{fig:fig_constant},~\ref{fig:fig_gaussian},~\ref{fig:fig_general} (while the posterior distributions and contours can be found in Appendix~\ref{sec:MCMC}). Each figure shows the correlation matrix used for the fitting, $\mathcal{F}$ (top-left panel), the model resulting from the best fit, $\mathcal{M}$ (top-right panel), the residuals between the two, $\mathcal{F}-\mathcal{M}$ (bottom-left panel), and the signal-to-noise ratio, $(\mathcal{F-M})/\sigma_\mathcal{D}$ (bottom-right panel). In the top-left panel, isocontours corresponding to the difference of the comoving distances at each redshift pair (i.e.~$\chi(z_1)-\chi(z_2)$, in units of Mpc) are also shown. The functional form for each model, the best fit and the minimum $\chi^2$ are shown in Table~\ref{tab:models}.

\subsection{$\chi^2$ evaluation}
To evaluate the goodness of fit for each model and compare between models, we consider the $\chi^2$ distribution and its minimum, $\min\chi^2$. The number of degree of freedom is listed in Table~\ref{tab:models} and defined as 
\begin{equation}
    N_\mathrm{dof} = N_{\mathrm{dof},\mathcal{D}} - \Theta_{\mathcal{M}},
\end{equation}
where $N_{\mathrm{dof},\mathcal{D}} = 221$ represents the number of independent redshift bin pairs in the correlation matrix $\mathcal{D}$ and $\Theta_{\mathcal{M}}$ the number of free parameters in each model $\mathcal{M}$. The models we test have only a few free parameters ($\Theta_{\mathcal{M}} \leq 4$), ensuring that $N_{\mathrm{dof},\mathcal{D}} \gg \Theta_{\mathcal{M}}$. We quantify the significance of the improvement in the fit between models using the likelihood ratio test \citep{Wilks38}, where the difference in $\chi^2$ values between two models is asymptotically distributed as a $\chi^2$ distribution with degrees of freedom equal to the difference of the models degrees of freedom. For example, a $\Delta\chi^2 = 10$ with $\Delta N_\mathrm{dof} = 2$ (1) degree of freedom corresponds to an improvement in the fit of a 2.5$\sigma$ (2.9$\sigma$) significance.

We also make use of the reduced $\chi^2$,
\begin{equation}
    \chi^2_{N_\mathrm{dof}} = \chi^2/N_\mathrm{dof},
\end{equation}
and its minimum $\min\chi^2_{N_\mathrm{dof}}$, which accounts for the number of degrees of freedom and is expected to be close to unity for a statistically consistent model under Gaussian errors, to evaluate the goodness-of-fit.

\section{Data fitting via models $\mathcal{M}$} \label{sec:results}

We tested our data against three functional forms of increasing complexity (Section~\ref{sec:functional_forms}), designed to capture the interplay between different scales. We present our findings in the following. A summary of the models definition, the best fits obtained via the MCMC and the minimum $\chi^2$, is listed in Table~\ref{tab:models}. \\ \\

The simplest model, $\mathcal{M}1$, assumes no scale dependent correlation and, as expected, provides the poorest fit to the data ($\chi^{2\,\,\mathcal{M}1}_{N_{\mathrm{dof}}} = 1.35$). In particular, it fails to reproduce the observed correlation between nearby redshift bins, which gradually decreases with separation at low redshift while strengthening at higher redshift. Nonetheless, the amplitude constrained by fitting the model to the data, $A = 0.09^{+0.01}_{-0.01} >0 $, seems to suggest a mild but significant overall correlation.

\begin{figure}
    \centering
    \centerline{\includegraphics[width=\columnwidth]{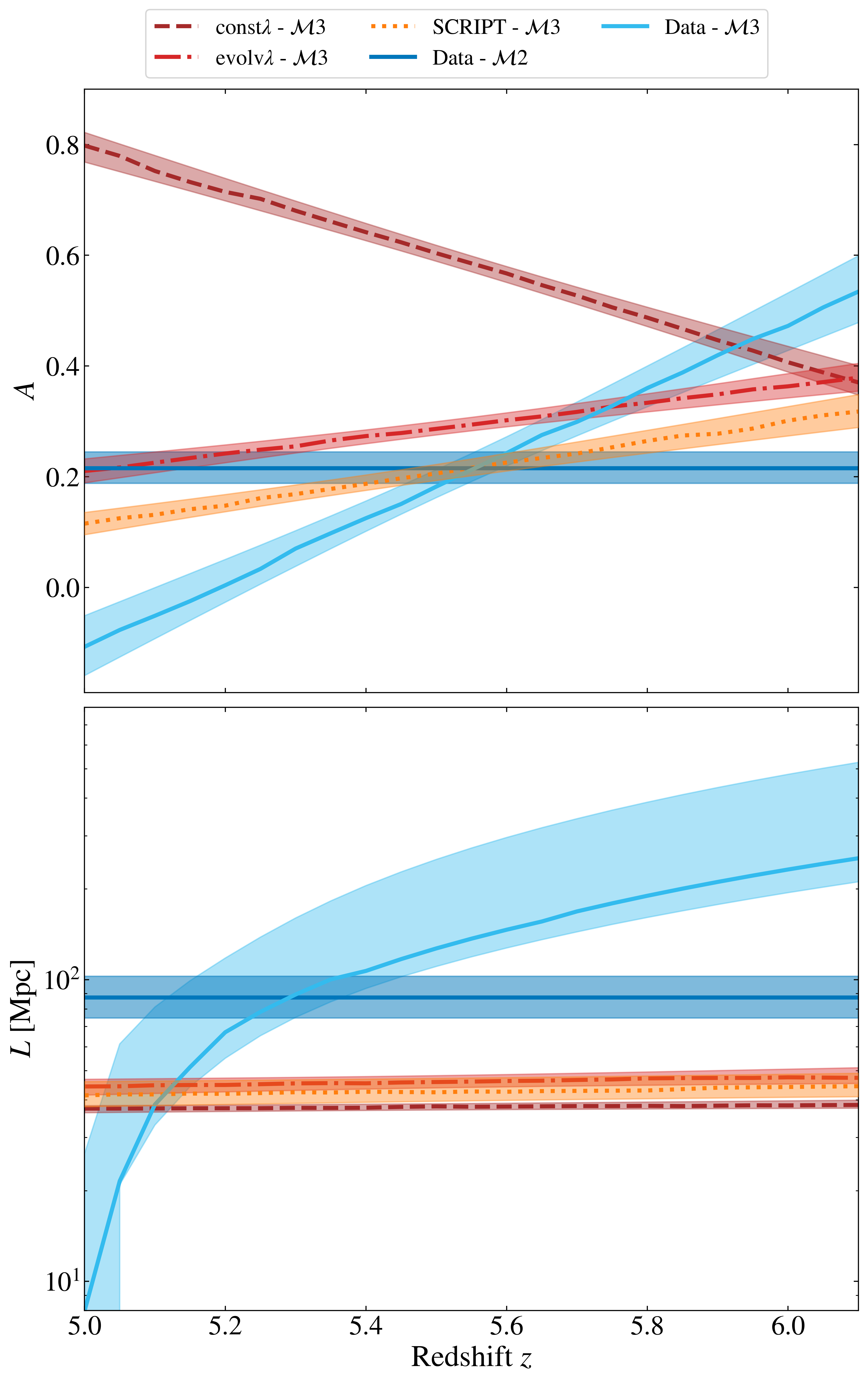}}
    \caption{Amplitude (top panel) and correlation length (bottom panel) as a function of redshift for different models and simulations. The continuous gray (blue) lines are the results of fitting the observations with $\mathcal{M}2$ ($\mathcal{M}3$), a Gaussian model with constant (redshift-dependent) width and amplitude. The gray shaded regions represent the $1-\sigma$ error on the $A$ ($L$) parameters. The blue shaded region instead indicates the 16th and 84th percentiles of the  $A_0$-$A_1$ ($L_0$-$L_1$) joint-posterior distribution, except for redshifts $z < 5.1$, where only the $1\sigma$ upper limit of $L$ is shown. The brown dashed, the red dash-dotted and the orange dotted lines are the results of fitting the simulations ($\mathrm{const}\lambda$, $\mathrm{evolv}\lambda$, $\mathrm{SCRIPT}$) with $\mathcal{M}3$.
    }
    \label{fig:fig_tot}
\end{figure}

Model $\mathcal{M}2$ assumes a strong correlation between nearby redshift bins that decays with separation. It is implemented as a Gaussian function centered on the diagonal of the correlation matrix, parameterized by the comoving separation of each redshift pair, with a constant amplitude and width. This model successfully captures the extent of correlation at low redshifts (see the bottom-left panel of Figure~\ref{fig:fig_gaussian}, which shows the residuals between the data and the model) but fails to reproduce the increasing strength of correlation at higher redshifts. The parameter $L$ in the functional form adopted (see Table~\ref{tab:models}) represents the characteristic correlation length. The best-fit values obtained via MCMC are $A = 0.22^{+0.03}_{-0.03}$ and $L = 86.16^{+16.42}_{-11.52}\,\mathrm{Mpc}$, indicating correlations between scales around 90 Mpc apart. The correlation length obtained for this model is also shown in Figure~\ref{fig:fig_tot} as a continuous gray line.
The comparison of the minimum $\chi^2$ values between the constant model $\mathcal{M}1$ and the Gaussian model with constant amplitude and width $\mathcal{M}2$ strongly favors the latter ($\Delta\chi^2 = \chi^{2}_{\mathcal{M}1} - \chi^{2}_{\mathcal{M}2} = 41.8$, with an improvement of $6.4\sigma$ significance). The improvement in the fitting is marked by a reduced $\chi^2$ of $\chi^{2\,\,\mathcal{M}2}_{N_{\mathrm{dof}}} = 1.17$ ($\chi^{2\,\,\mathcal{M}1}_{N_{\mathrm{dof}}} = 1.35$). 

Only our last model $\mathcal{M}3$, Gaussian with a redshift-dependent amplitude and width, is able to capture the correlation at high redshifts and allow us to provide an estimate of its strength and length. For a fixed redshift $z_1$, the Gaussian profile evolves with $z_2$ through its amplitude and width, both of which are allowed to vary linearly with $z_2$ (see Table~\ref{tab:models}). This additional flexibility enables $\mathcal{M}3$ to better describe the evolution of correlations across redshifts. The residuals, shown in the bottom-left panel of Figure~\ref{fig:fig_general}, demonstrate a significantly improved fit, particularly at high redshift. The posterior distribution (see Figure~\ref{fig:MCMC_general} in Appendix~\ref{sec:MCMC}) shows a well constrained amplitude of $A = -0.10^{+0.05}_{-0.05} + 0.58^{+0.10}_{-0.09} \cdot (z_2-5)$, mildly evolving with redshift. The width however, defined as $L(z_1,z_2) = L_0+L_1\cdot(z_2-5) $, is less tightly constrained: while its evolution with redshift is well recovered in the MCMC run (through the $L_1$ parameter), its constant counterpart ($L_0$) is constrained only as an upper limit. The best-fit parameters read $L_0 \leq 76.45\,\mathrm{Mpc}$ at the $2\sigma$ level and $L_1 = 283.77^{+176.24}_{-117.45}\, \mathrm{Mpc}$, suggesting a combination of a general, overall, correlation with a strong, redshift-dependent one. The correlation scale $L$ as a function of redshift is shown in Figure~\ref{fig:fig_tot} as a continuous blue curve, representing the maximum likelihood from the joint posterior distribution of the $L_0$ and $L_1$ parameters at each redshift. The shaded region indicates the 16th and 84th percentiles of these distributions, except for redshifts $z < 5.1$, where only an upper limit of $L$ is shown. Our results suggest a redshift-dependent correlation length, from $L \leq 26.53\, (68.47)\,\mathrm{Mpc}$ at 1-$\sigma$ (2-$\sigma$) at redshift $z = 5.0$ to $L = 252.72^{+272.61}_{-41.61}\,\mathrm{Mpc}$ at redshift $z = 6.1$.

The best-fit $\chi^2$ obtained from the fitting procedure is 202.9, corresponding to a reduced $\chi^2$ of $\chi^{2\,\,\mathcal{M}3}_{N_{\mathrm{dof}}} = 0.94\sim 1$, strongly favoring $\mathcal{M}3$ over $\mathcal{M}2$ ($\Delta\chi^2 = \chi^{2}_{\mathcal{M}2} - \chi^{2}_{\mathcal{M}3} = 53.7$, with an improvement of $7\sigma$ significance). 

In summary, all models detect  correlations in the Lyman-$\alpha$ forest flux across redshift, with their inferred strength and scale depending on the model. The best-fit $\chi^2$ (and reduced $\chi^2$) is achieved with our most general function, $\mathcal{M}3$, which includes a redshift-dependent correlation length, suggesting coherent structures extending over $\gtrsim 200$ Mpc at the highest redshifts $z>6$. The resulting best-fit correlation scales as a function of redshift are shown in Figure~\ref{fig:fig_tot}.

Finally, in order to be as conservative as possible, we repeat the analysis while allowing for a possible under-estimation of the uncertainties arising from continuum flux reconstruction, which are correlated across all scales. To do this, we 
perturb each quasar in the mock sample by a factor $1+ \mathcal{G}(1.5\cdot\sigma_{z, \,\mathrm{QSO}}^\mathrm{cont})$, i.e.~with a increase of $50\%$ in the uncertainties compared to the analysis above, and far more pessimistic than the accuracy measurements for the PCA-based reconstruction method we are employing. The results 
are shown in Figure~\ref{fig:fig_tot_1.5sigma}. 
The measurement of $L(z)$ when increasing the continuum-reconstruction error by 50\% are consistent within 10\% with the fiducial analysis above.

\section{Simulations} \label{sec:simulations}

\begin{figure*}
    \centering
    \centerline{\includegraphics[width=\textwidth]{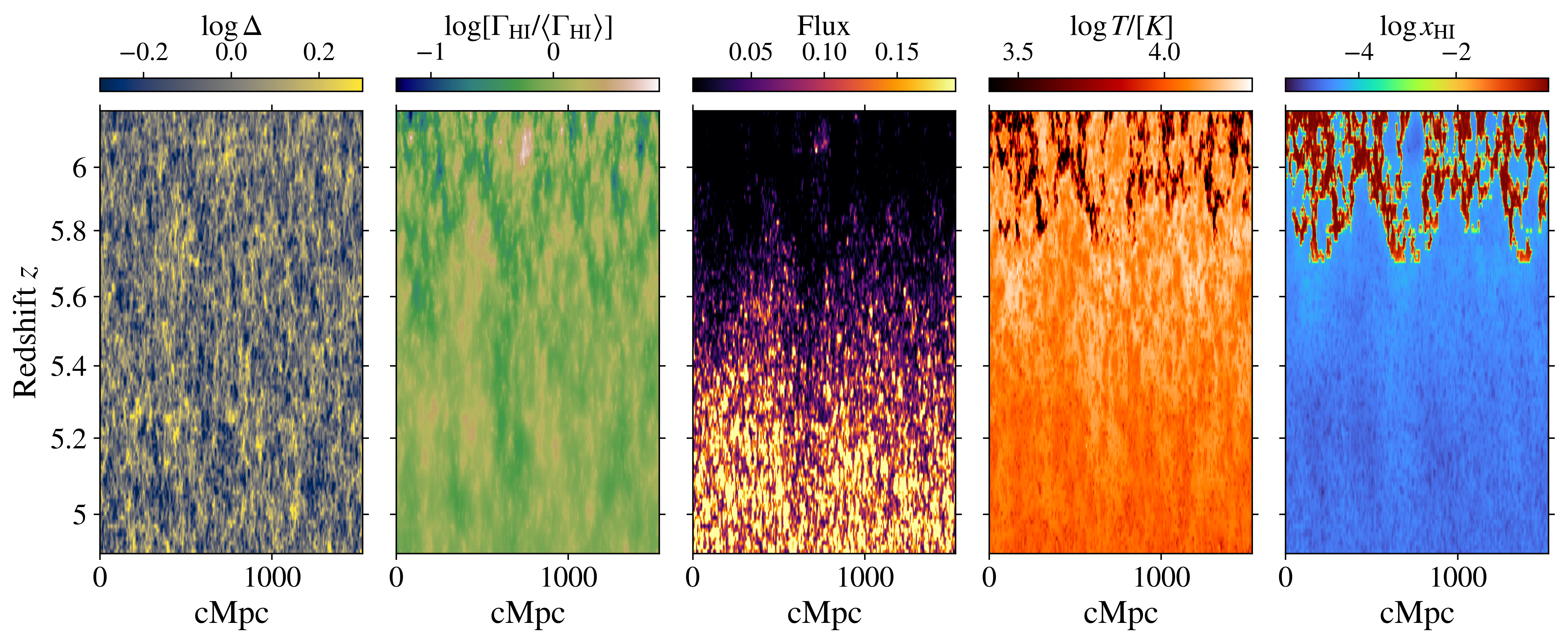}}
    \caption{Lightcone from BIBORTON simulations, with SCRIPT model ($\mathrm{SCRIPT}$, see Section~\ref{sec:simulations}). From left to right, we show the underlying matter density, the normalized photo-ionization rate, the flux, the temperature and the HI fraction. The lightcone fully covers the redshift range considered in this work and has a size of $\sim 1.5$ cGpc.}
    \label{fig:script}
\end{figure*}

To gain insights into the physical processes occurring during the late EoR, we wish  to compare our observational findings with physically-motivated IGM simulations. For this study, we would ideally require a simulation box with a high dynamic range that preserves large-scale correlations while simultaneously capturing the small scale physics. However, that is unfortunately computationally intractable with currently available resources. Instead, we use a complementary method and employ an optimized approach using a semi-numerical technique developed by \citet{Maity25}. This method can efficiently generate large-volume lightcones of Ly-$\alpha$ transmission across the redshift range ($z=4.9-6.2$, separated by a comoving length of $\sim 420 \,h^{-1}\mathrm{Mpc}$) relevant to our study. Below, we briefly summarize the simulation procedures of the BIBORTON (BIg BOxes for ReionizaTiON) suite of simulations, that will be fully described in Maity et al.~(in prep).

In the semi-numerical setup, the Ly-$\alpha$ optical depth is assumed to depend on underlying cosmological density fluctuations, ultraviolet background (UVB) fluctuations, temperature fluctuations, and ionization fluctuations (if reionization is incomplete). These dependencies are calibrated against a high resolution full hydrodynamic simulation \citep[in this case \texttt{Nyx},][]{Almgren13}. Additionally, the semi-numerical reionization and UVB generation models require information on the collapsed halo mass fraction ($f_{\mathrm{coll}}$) to quantify available ionizing sources. For the purpose of this work, we generate density fields with a large box size ($L$) of $1024 ~h^{-1}\mathrm{Mpc}$ at a given redshift using the Zel’Dovich Approximation \citep{zeldovich70} while the collapsed fraction field is generated using Excursion Set Formalism in Lagrangian space \citep[ESF-L;][]{Trac22}.
This provides sufficient resemblance with full N-body simulation at large scales while allowing us to efficiently investigate the EoR evolution . 
We work with a spatial resolution ($\Delta x$) of $4~h^{-1}\mathrm{Mpc}$ allowing an optimization between efficiency and accuracy. To get the lightcone evolution, we generate the boxes at a redshift interval, $\Delta z=0.1$ and interpolate the redshifts in-between. Given the density and source information, we utilize the EX-CITE model \citep{Gaikwad23} to efficiently generate UVB fluctuations while incorporating local source contributions following \citet{Davies16,Davies24}. This approach requires two input parameters: the effective mean free path of ionizing photons ($\lambda_\mathrm{mfp}$) and the mean photoionization rate ($\langle\Gamma_{\mathrm{HI}}\rangle$). Furthermore, ionization and temperature evolution are modeled using the photon-conserving reionization model, SCRIPT, which accounts for recombination effects \citep{Maity22}. In this study, we do not explicitly include radiative feedback for computational convenience; instead, we account for its effects by imposing a constant threshold halo mass ($M_{\mathrm{min}}=10^9 M_{\odot}$), below which collapsed objects cannot form. 
We explore three different scenarios which are listed below:

\begin{itemize}

\item \underline{$\mathrm{const}\lambda$: BIBORTON with constant $\lambda_\mathrm{mfp}$}: In this case, we assume a constant mean free path \citep[$\lambda_\mathrm{mfp}=8~h^{-1}\mathrm{Mpc}$,][]{Maity25} throughout the redshift range for UVB generation. We further assume that the reionization is over at redshift $z \geq 6.1$ (i.e.~no neutral island is left in the redshift range considered in this work) and the temperature is estimated using the standard power law relation with density, $T=T_0\Delta^{\gamma-1}$ \citep[with $T_0=10^4~ K$ and $\gamma=1.35$, ][]{Maity25}. \\

\item \underline{$\mathrm{evolv}\lambda$: BIBORTON with evolving $\lambda_\mathrm{mfp}$}: This is similar to the previous model except we now allow for redshift-dependent evolution of $\lambda_\mathrm{mfp}$. Specifically, we assume a linear variation from $4~h^{-1}\mathrm{Mpc}$ at $z=6.2$ to $25 ~h^{-1}\mathrm{Mpc}$ at $z=4.9$. 
\\

\item \underline{$\mathrm{SCRIPT}$: BIBORTON with SCRIPT} (\textit{fiducial}): This case incorporates a realistic reionization scenario (with reionization end at $z\approx 5.7$ and neutral island present until such moment) generated with SCRIPT, as described earlier. This model now incorporates self consistent temperature fluctuations satisfying IGM temperature estimates \citep{Gaikwad20}. The evolution of $\lambda_\mathrm{mfp}$ is assumed to be same as the previous one. The underlying matter density, the normalized photo-ionization rate, the flux, the temperature evolution and the HI fraction for this model are shown in Figure~\ref{fig:script}.

\end{itemize}

In each of these cases, we tune the mean photoionization rate at each redshift to match the observed distribution of mean transmission fluxes \citep{Bosman22}.

Once all the necessary fields are generated, we then construct the lightcone volume for the redshift range studied in this work. Notably, the lightcones are not affected by any correlations coming from boundary discontinuities as our box size is large enough to cover the whole comoving length corresponding to the redshift range of our interest. Finally, we extract 1000 random skewers along the line of sight which are utilized for correlation scale analysis. 

In order to properly compare the random skewers extracted from the simulations with the observations, observational noise must be incorporated into the former. We achieve this by perturbing each simulated skewer at each redshift. Specifically, for each skewer, we randomly select a quasar from the observed sample and, at each redshift, add a random noise component drawn from a Gaussian distribution $\mathcal{G}(\sigma_{z, \,\mathrm{QSO}}^\mathrm{obs})$ with a standard deviation $\sigma_{z, \,\mathrm{QSO}}^\mathrm{obs}$ equal to the observational uncertainty of the chosen quasar at that redshift. Given that the observational uncertainties are not correlated across redshift, each redshift in the simulated skewer is perturbed independently using the corresponding uncertainty of the selected quasar. Since the continuum-reconstruction uncertainties are accounted for at the observations level by subtracting their contribution from the observed correlation matrix (Section~\ref{sec:methods}), we do not correct the simulations for the continuum-reconstruction procedure. This procedure ensures that the simulated skewers closely mimic the statistical properties of real observations while preserving the underlying large-scale correlations.

\begin{figure}
    \centering
    \centerline{\includegraphics[width=\columnwidth]{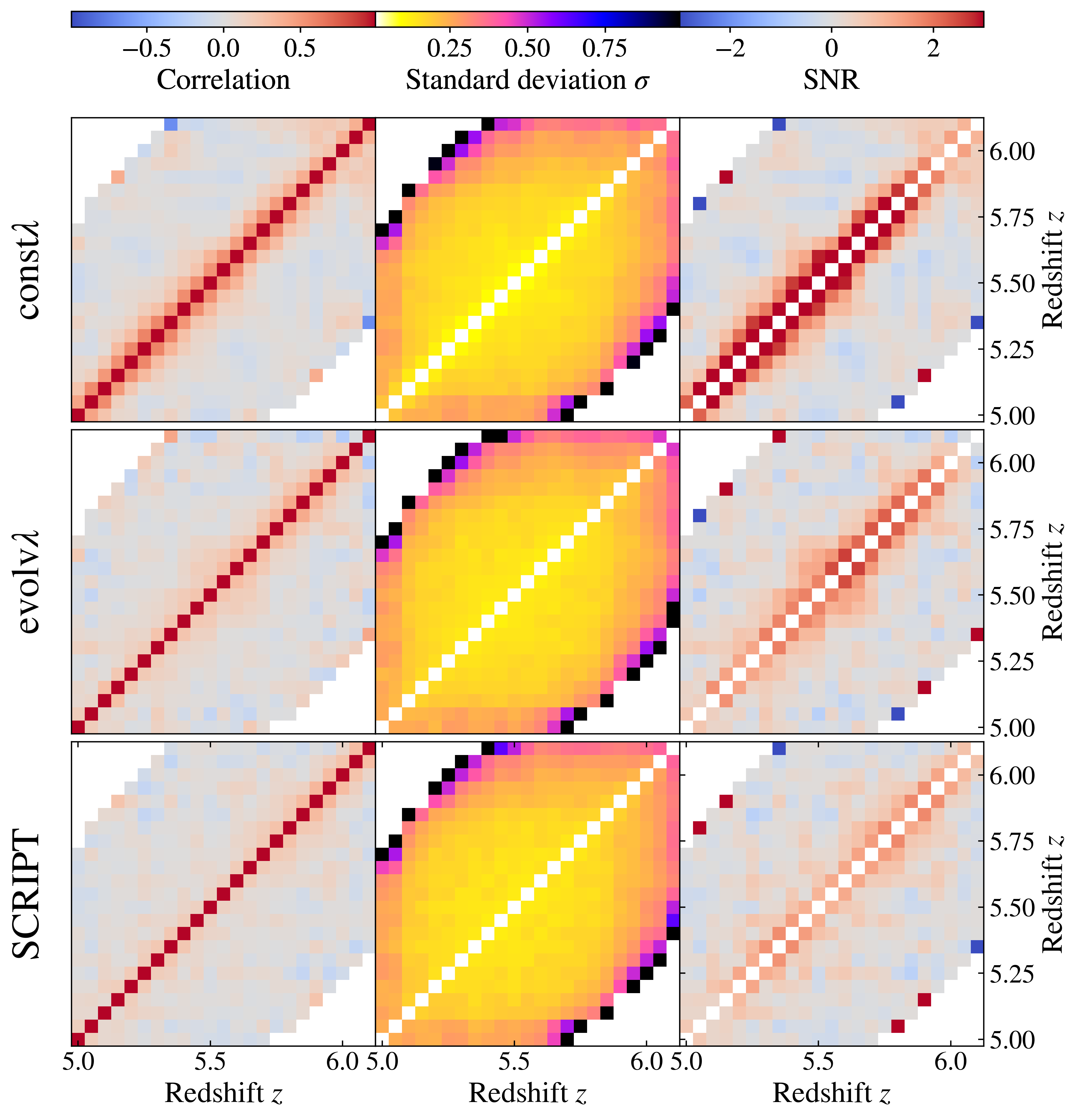}}
    \caption{As Figure~\ref{fig:fig2} for the BIBORTON simulations. Note the different range of colors for second and third columns.}
    \label{fig:fig_BBBB}
\end{figure}

\subsection{Correlation matrix}

We compute the correlation matrix for the 1000 random skewers extracted from lightcones in the BIBORTON simulation suite. The resulting correlation matrices, their standard deviations (each obtained by building $N = 1000$ correlation matrices of 67 random sightlines tailored to reproduce the observational sample), and their significance, are shown in the columns of Figure~\ref{fig:fig_BBBB}. Each row represents one of the lightcones of the simulation suite ($\mathrm{const}\lambda$, $\mathrm{evolv}\lambda$, $\mathrm{SCRIPT}$). Interestingly, the correlation between redshift bins exhibit in the lightcones is consistently weaker than the one observed in the data. 
A mild correlation between nearby redshift bins is visible for a constant $\lambda_\mathrm{mfp} = 11.94\, \mathrm{Mpc}$ ($\mathrm{const}\lambda$), both at high and low redshift, slightly increasing towards higher redshifts. In $\mathrm{evolv}\lambda$, the mean free path evolves from $\lambda_\mathrm{mfp} = 5.97 \, \mathrm{Mpc}$ at $z = 6.2$ to $\lambda_\mathrm{mfp} = 37.31 \, \mathrm{Mpc}$ at $z = 4.9$ (i.e.~$\lambda_\mathrm{mfp}(z = 5.95) = 11.94 \, \mathrm{Mpc}$); the correlation is lower then for $\mathrm{const}\lambda$ at high redshift, but remains comparable at low ones (where the mean free path is approximately triple), in line with the mild correlation at low redshift observed in the data. The last simulation, $\mathrm{SCRIPT}$ has the same mean free path evolution of $\mathrm{evolv}\lambda$ but later ending of reionization ($z \approx 5.7$); a broader, but weaker with respect to $\mathrm{const}\lambda$ and $\mathrm{evolv}\lambda$, correlation is detected at all redshifts.

\subsection{Modeling through BIBORTON simulations} \label{sec:model_BIBORTON}
The simple functional forms presented in Section~\ref{sec:functional_forms} provide a clear interpretation and measure of the correlation length, either through the parameter $L$ or its absence. Once such a correlation has been estimated, we aim to explore the underlying physical processes responsible for setting the observed correlation scale. For this reason, we compare the observations with the BIBORTON simulations, each representing a different model for the late stages of the EoR. In particular, both $\mathrm{const}\lambda$ and $\mathrm{evolv}\lambda$ assume an early-ending reionization scenario ($z \geq 6.1$), with $\mathrm{const}\lambda$ adopting a constant $\lambda_\mathrm{mfp}$ and $\mathrm{evolv}\lambda$ allowing it to evolve with time. Meanwhile, $\mathrm{SCRIPT}$ incorporates a more realistic reionization history, ending at $z \approx 5.7$ (with neutral islands present until then). The results are presented in Figure~\ref{fig:fit_S3} for our fiducial simulation $\mathrm{SCRIPT}$, in Figures~\ref{fig:fit_S1} and~\ref{fig:fit_S2} for the other models, and summarized in Table~\ref{tab:models}. 

None of the simulations, however, successfully reproduce the observed correlations. While the behavior at low redshift is generally recovered (the residuals normalized by the standard deviation are in general consistent with zero and randomly distributed), the residual panels in each figure reveal a lack of correlation strength at high redshift. The poor fit is further confirmed by the $\chi^2$ and reduced $\chi^2_{N_\mathrm{dof}}$ values, which yield for $\mathrm{const}\lambda$ and $\mathrm{evolv}\lambda$ values of $\chi^{2\,\mathrm{const}\lambda}_{N_\mathrm{dof}}= 392.3/221 = 1.76$ and $\chi^{2\,\mathrm{evolv}\lambda}_{N_\mathrm{dof}}= 278.8/221 = 1.26$.
The fiducial simulation $\mathrm{SCRIPT}$ shows the best agreement with observations ($\chi^{2\,\mathrm{SCRIPT}}_{N_\mathrm{dof}}= 274.6/221 = 1.24$), presenting a broad but weak correlation at all scales (but failing to reproduce the high redshift one). This is despite the fact that all three simulations match the observed distribution of Lyman-$\alpha$ transmission at these redshifts, meaning that high optical depth in the forest is not sufficient to reproduce the correlations we observed.

We test how our most general model, $\mathcal{M}3$, fit the simulations. As a Gaussian model with a redshift-dependent amplitude and width, $\mathcal{M}3$ can, in principle, reproduce both $\mathcal{M}1$ (which corresponds to no correlation, i.e., $L \sim 0$ Mpc or considerably small) and $\mathcal{M}2$ (which assumes a redshift-independent correlation, i.e., $L_1 \sim 0$ Mpc). We present the results of this fit in Figures~\ref{fig:fit_bbbb_with_M3},~\ref{fig:mcmc_bbbb_with_M3} and Table~\ref{tab:models}, while the evolution of the correlation length with redshift is shown in Figure~\ref{fig:fig_tot}. 

The model captures the main features of the lightcones, yielding minimum $\chi^2$ values that are generally higher than those obtained for the observational fits, though still comparable ($\chi^{2\,\mathrm{const}\lambda-\mathcal{M}3}_{N_\mathrm{dof}} = 1.23$, $\chi^{2\,\mathrm{evolv}\lambda-\mathcal{M}3}_{N_\mathrm{dof}} = 1.45$, $\chi^{2\,\mathrm{SCRIPT}-\mathcal{M}3}_{N_\mathrm{dof}} = 1.36$). 

The constraints on the free parameters differ significantly from those obtained in the observations. For instance, the redshift-dependent component of the amplitude, $A_1$, is now constrained to be negative for $\mathrm{const}\lambda$ ($A_1 = -0.36^{+0.02}_{-0.06}$), leading to an overall decreasing amplitude with redshift. The amplitude parameter $A$ recovered for each simulation and when fitting the data is shown in the top panel of Figure~\ref{fig:fig_tot}. The amplitude recovered when fitting the data with $\mathcal{M}3$, strongly increases with redshift, while it has a milder evolution for $\mathrm{const}\lambda$ and $\mathrm{evolv}\lambda$ simulations.

The main difference, however, concerns the correlation length: a weak redshift dependence is now preferred over a strong one, with $L_1$
consistent with zero in all cases. The correlation length recovered in each simulation is generally around $\sim50 \,\mathrm{Mpc}$, much lower than the one recovered by fitting the data with the same model (spanning from $50 \,\mathrm{Mpc}$ at low redshift to $250 \,\mathrm{Mpc}$ at high redshift). The amplitude and correlation lengths constrained with the simulations, as a function of redshift are shown in Figure~\ref{fig:fig_tot}, as the brown dashed line ($\mathrm{const}\lambda$), the red dash-dotted line ($\mathrm{evolv}\lambda$) and the orange dotted line ($\mathrm{SCRIPT}$). The shaded regions are computed as the 16th and 84th percentiles of the $A_0$-$A_1$ and $L_0$-$L_1$ joint-posterior distribution. 

For $\mathrm{const}\lambda$, the simulation with constant mean free path, the correlation length suggested by fitting $\mathcal{M}3$ has a very weak dependence on redshift, going from $L = 37.39^{+0.89}_{-1.13} \,\mathrm{Mpc}$ at redshift $z = 5.0$ to $L = 38.45^{+1.45}_{-0.91}\,\mathrm{Mpc}$ at redshift $z = 6.1$. A strong (SNR $\geq 3$) correlation is detected between nearby redshift bins, with a scale of $\sim 3$ times the constant mean free path implemented in the simulation. 

When allowing the mean free path to vary with redshift (spanning from $\sim 6$ Mpc to $\sim 37$ Mpc), as for the simulation $\mathrm{evolv}\lambda$, the correlation length results higher than for $\mathrm{const}\lambda$, spanning from $L = 44.27^{+2.53}_{-2.79} \,\mathrm{Mpc}$ at redshift $z = 5.0$ to $L = 47.29^{+3.83}_{-1.96}\,\mathrm{Mpc}$ at redshift $z = 6.1$. Correlations with a SNR $\geq 2$ are strongly detected between nearby redshift bins but only in the central redshift interval ($\sim 5.3-5.8$). The short mean free path implemented at high redshift does not allow the ionizing photons to causally connected more distant regions.

Finally, postponing the end of the EoR to $z \approx 5.7$ (i.e.~introducing neutral islands) has a similar effect, leading to a correlation length of $L = 41.37^{+4.61}_{-3.44} \,\mathrm{Mpc}$ at redshift $z = 5.0$ to $L = 44.35^{+4.64}_{-3.38}\,\mathrm{Mpc}$ at redshift $z = 6.1$.

\section{Summary \& Conclusion} \label{sec:discussion}

The existence of Gunn-Peterson troughs over 100 Mpc in length at $z\geq5.7$ signaled the fact the ionisation state of the IGM is potentially correlated on scales far larger than the density field \citep{Becker15}. We used 67 Lyman-$\alpha$ forest skewers from the E-XQR-30 dataset to quantify these correlation scales at $5.0<z<6.1$ and investigate potential redshift evolution. 

The normalised covariance matrix of the Lyman-$\alpha$ flux, computed in bins of $\Delta z=0.05$, is shown in Figure \ref{fig:fig2}. After carefully accounting for systematics as described in Section~\ref{sec:methods}, we find statistically significant correlations on extremely large scales: the transmitted flux at $z=5.6$ is positively correlated with all higher redshifts along the same skewers. 

We fit the correlation matrix and its uncertainties with a function describing a redshift-independent gaussian correlation in redshift ($\mathcal{M}2$), finding a best-fit characteristic scale of $L = 86.16_{-11.52}^{+16.42}$ Mpc. This simple model is strongly preferred over a model with no correlation between redshifts ($\mathcal{M}1$): the minimum (reduced) $\chi^2$ drops from $298.4$ to $256.6$ ($1.35$ to $1.17$) with the addition of $1$ extra parameter. 

Next, we allow the gaussian correlation scale to vary linearly with redshift ($\mathcal{M}3$), finding again a statistically significant improvement with the minimum (reduced) $\chi^2$ dropping to $202.9$ ($0.94$) by adding $2$ free parameters. The recovered characteristic correlation scales increase from $L< 76.45$ Mpc at $z=5.0$ ($2\sigma$ limit) to $L=252.72_{-41.61}^{+272.61}$ Mpc at $z=6.1$. In other words, at $z=6.1$, the Lyman-$\alpha$ forest is correlated on scales no smaller than $200$ Mpc: far larger than typical numerical (and semi-numerical) models of the reionisation process.

To explore whether semi-numerical models of the density field and reionisation are able to reproduce correlations on such scales, we use the recent BIBORTON simulation framework \citep{Maity25}. This suite uses a simulation of the density field in a very large box, 1024 $h^{-1}$ Mpc on the side, onto which UVB fluctuations are added using EX-CITE \citep{Gaikwad23}. We generate lightcones spanning the same redshift range as the observations, forward-model Lyman-$\alpha$ forest skewers, and compute the covariance matrix in the same way as for the observed dataset. We find that three versions of BIBORTON, assuming either a constant mean free path, an evolving mean free path, or fully processed through SCRIPT, all display significantly shorter correlation scales than the observations and provide a worse fit than our simple assumption of a redshift-independent gaussian correlation ($\mathcal{M}2$) Despite the fact that they match the observed mean flux and scatter of the Lyman-$\alpha$ forest, the BIBORTON simulations all have characteristic correlation scales $<60$ cMpc at $z=6$, in strong tension with our observations, \textit{by a factor of over four}.

A correlation scale in excess of 100 Mpc is perhaps not surprising in the context of the existence of large Gunn-Peterson troughs at $z>5.5$ \citep{Becker15}; however, such troughs are still rare at $z<6$ \citep{Zhu21}, so additional correlations within the transmissive regions of the IGM may still be required to explain our measurement. A hint of the excess correlations on large-scales can be seen in auto-correlation function measurements of \citet{Wolfson23}: while they only considered scales up to $L\sim15$ cMpc (the size of our individual flux bins), they generally found indications for elevated large-scale correlations compared to simulations. 

A potential explanation for these discrepancies lies in the reionization models implemented in the simulations. Both $\mathrm{const}\lambda$ and $\mathrm{evolv}\lambda$ assume an early and uniform end to reionization, with no surviving neutral islands at the redshifts of interest. While this could initially be considered a reason for the lack of correlation, $\mathrm{SCRIPT}$, which includes neutral islands persisting to lower redshifts, still does not match the observations. This suggests that the presence of UVB fluctuations and neutral structures alone do not explain the observed correlation trends.

Furthermore, an evolving mean free path of ionizing photons does not seem, by itself, to significantly affect correlation strength, as shown by the similarity in correlation behavior between $\mathrm{const}\lambda$ and $\mathrm{evolv}\lambda$. Interestingly, in $\mathrm{SCRIPT}$, where neutral islands persist, the correlation strength is lower than in $\mathrm{evolv}\lambda$ at redshifts where reionization is still ongoing. This suggests a complex relationship between the mean free path, neutral islands, and large-scale correlations.

Future $z>6$ quasars to be soon discovered by \textit{Euclid} \citep{Euclid25} and followed up by \textit{ELT} \cite{ELT-ANDES} will enable a more precise characterization of large-scale IGM correlations. Euclid’s wide-area coverage will increase the number of high-$z$ quasars, while the ELT will be able to deliver high-SNR spectra of these fainter sources. Potentially, the number of known quasars during the end of the EoR may one day be sufficiently large to enable the search for transverse correlations between sightlines in addition to the purely longitudinal ones considered in this work.

\section*{Data Availability}

The data presented in this article will be shared on reasonable request to the corresponding author (BS). The correlation matrix of the QSO sightlines computed and used in this work is available on the first author's website\footnote{\url{https://sites.google.com/view/benedettaspina/about/}}.

\begin{acknowledgements}
BS and SEIB are supported by the Deutsche Forschungsgemeinschaft (DFG) under Emmy Noether grant number BO 5771/1-1. Part of this work is based on observations collected at the European
Southern Observatory under ESO programme 1103.A-0817. We thank Andrei Mesinger, Anna Pugno and Elena Marcuzzo for the useful comments and discussions. 
\end{acknowledgements}

\bibliographystyle{aa}
\bibliography{aanda}

\clearpage

\begin{appendix}

%\vfill 

%\begin{minipage}{\textwidth}
\section{QSO sample} \label{sec:QSO_list}
\begin{figure}[H]
\makebox[\textwidth][c]{\includegraphics[width=0.85\textwidth]{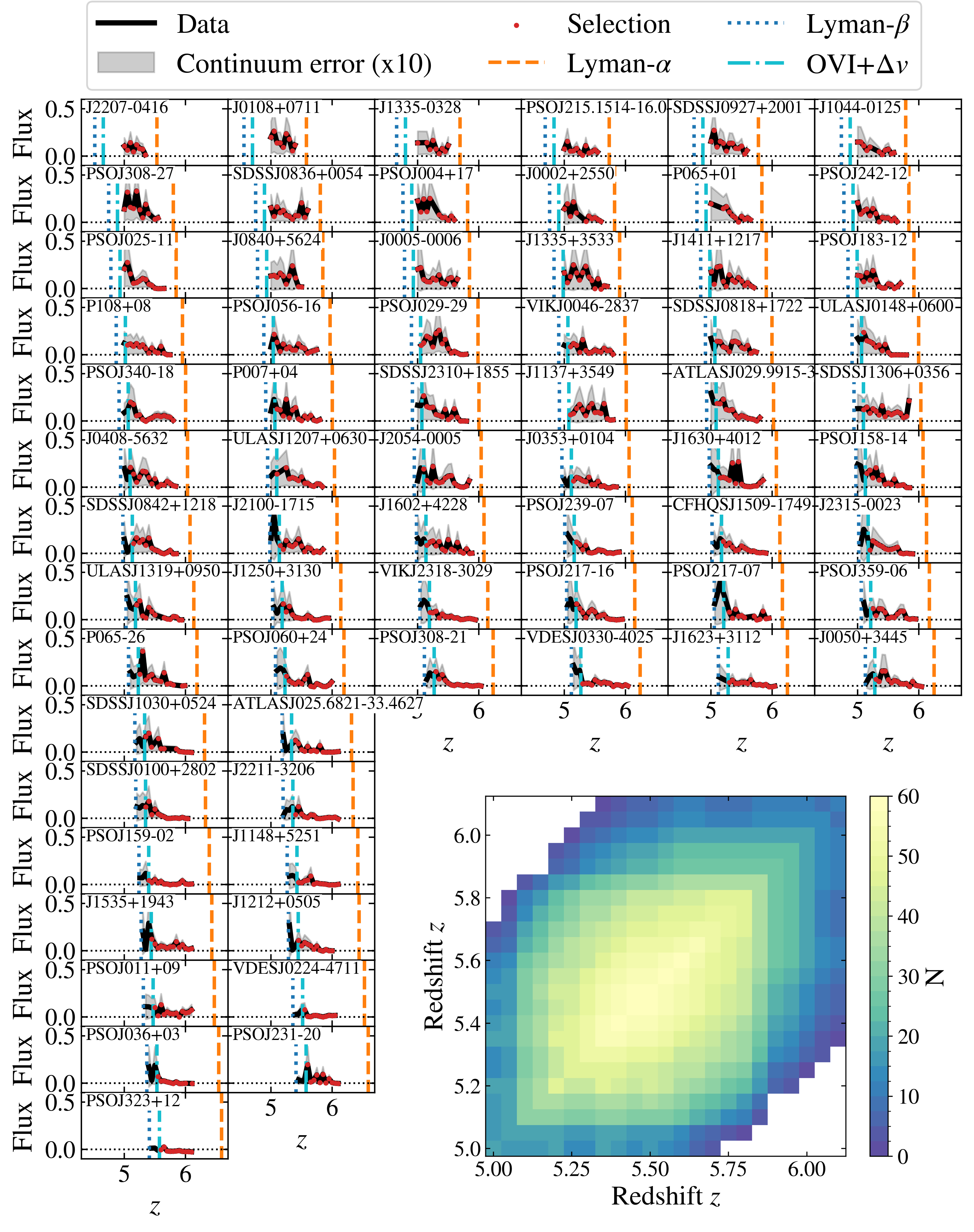}}
%\caption{The quasar sample, as described in Sec.~\ref{sec:sample}. Each quasar is displayed in a different panel, ordered by redshift, with a binning of $\Delta z = 0.05$. The detected flux as a function of redshift is shown as a black continuous line. The gray shaded region represents the continuum-reconstruction uncertainty. Data selected based on the OVI doublet emission line (semi-dashed light blue vertical line) are highlighted in red. The Lyman-$\alpha$ and Lyman-$\beta$ lines are indicated by the dashed orange and dotted blue lines, respectively. \bene{The number of sightlines available in each redshift bin (and their combination) is reported in the bottom-right panel.}}
%\label{fig:fig1}
\end{figure}

\begin{center}
    \parbox{0.95\textwidth}{
        Fig. A.1: The quasar sample, as described in Sec.~\ref{sec:sample}. Each quasar is displayed in a different panel, ordered by redshift, with a binning of $\Delta z = 0.05$. The detected flux as a function of redshift is shown as a black continuous line. The gray shaded region represents the continuum-reconstruction uncertainty. Data selected based on the OVI doublet emission line (semi-dashed light blue vertical line) are highlighted in red. The Lyman-$\alpha$ and Lyman-$\beta$ lines are indicated by the dashed orange and dotted blue lines, respectively. \bene{The number of sightlines available in each redshift bin (and their combination) is reported in the bottom-right panel.}
    }
\end{center}
%\end{minipage}

$ $ \\

$ $ \\

$ $ \\

$ $ \\

$ $ \\

\vspace{20cm}

\begin{minipage}{\textwidth}
\section{Fitting the data: models MCMC posteriors} \label{sec:MCMC}

\begin{figure}[H]
    \centering
    \makebox[\textwidth][l]{
        \subfloat[Constant model, $\mathcal{M}1$.\label{fig:MCMC_constant}]{\includegraphics[width=0.4\textwidth]{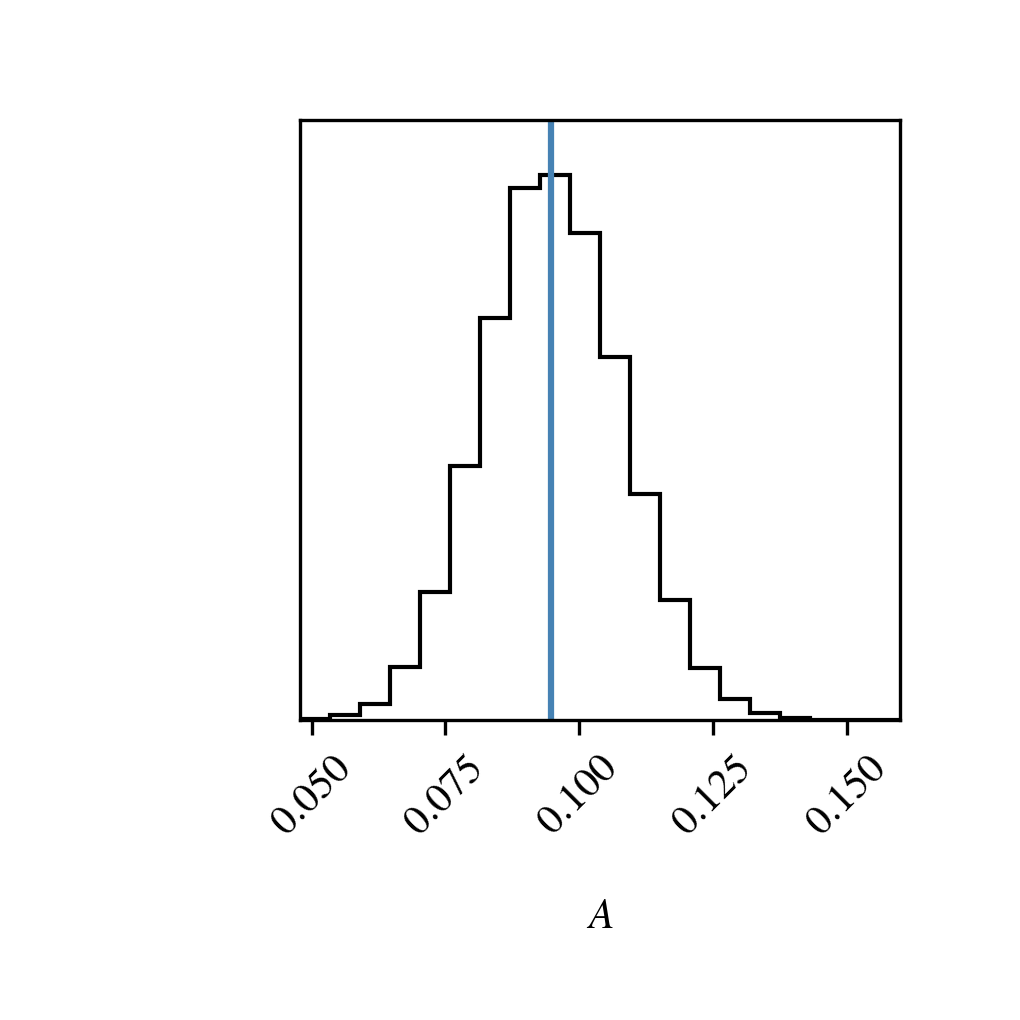}}
        \subfloat[Gaussian model with constant amplitude and width, $\mathcal{M}2$.\label{fig:MCMC_gaussian}]{\includegraphics[width=0.4\textwidth]{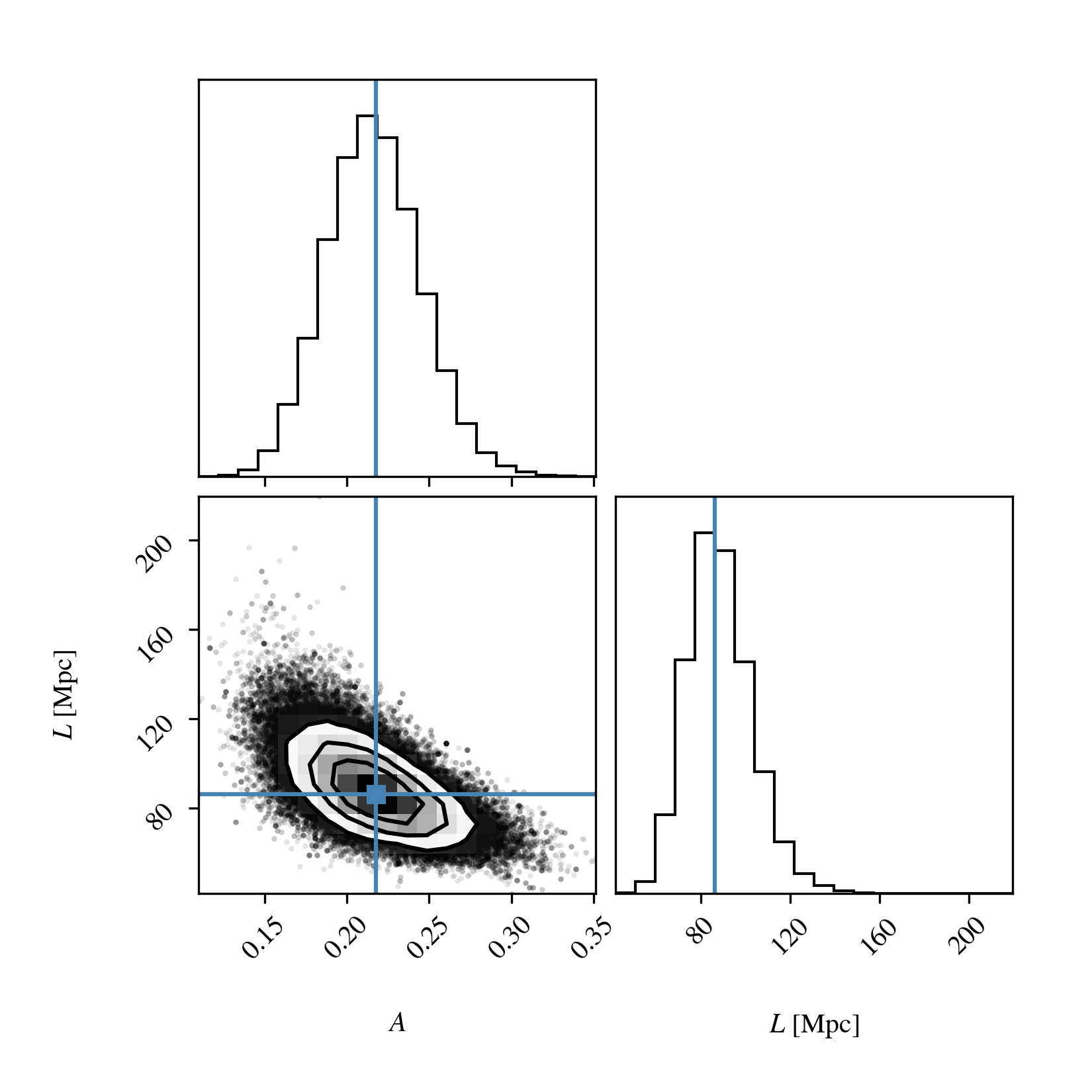}}
    }
    \makebox[\textwidth][c]{
        \subfloat[Gaussian model with redshift-dependent amplitude and width, $\mathcal{M}3$. For convenience, $L_0$ and $L_1$ are sampled logarithmically.\label{fig:MCMC_general}]{\includegraphics[width=0.7\textwidth]{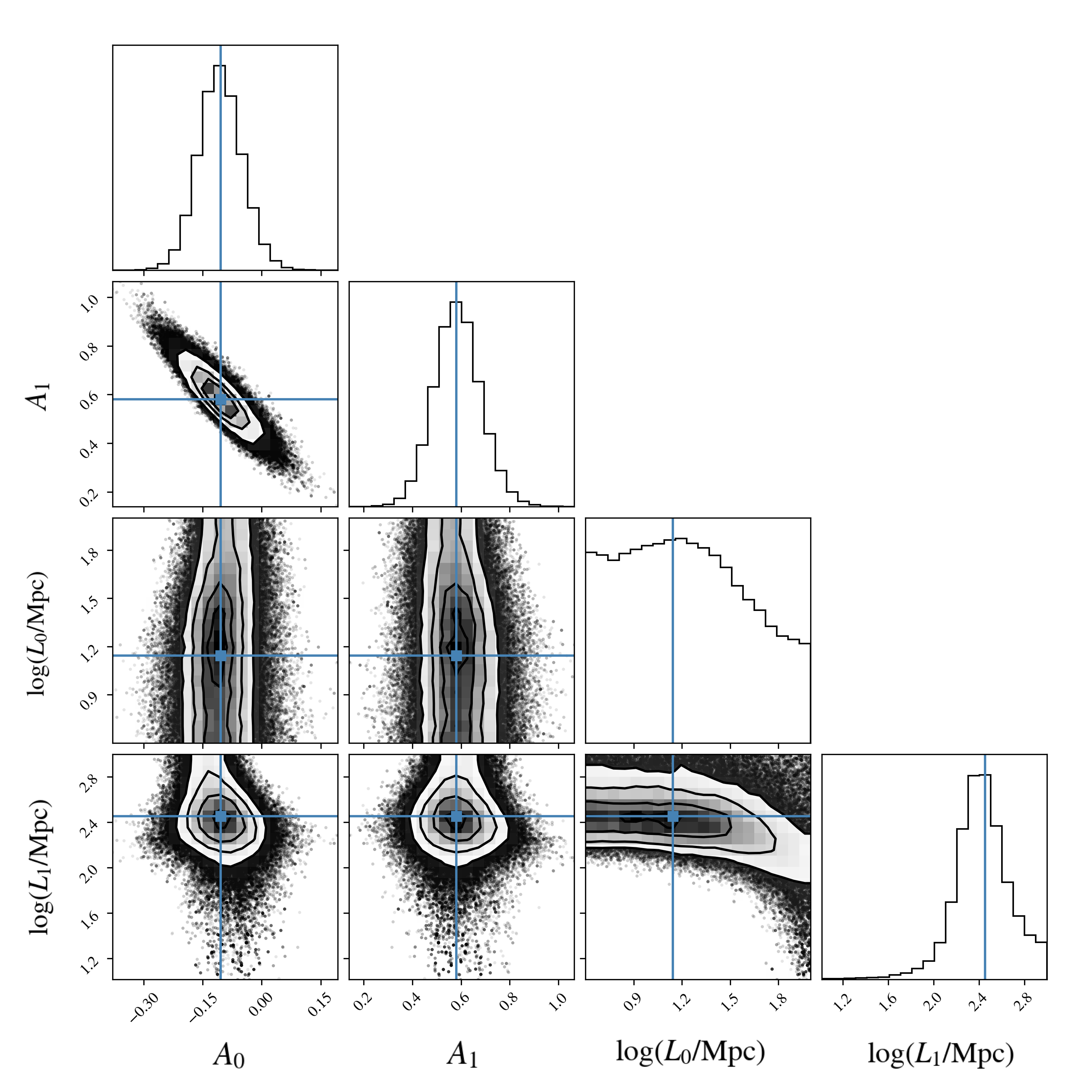}}
    }
    \caption{Posterior and best fit from the MCMC for the different models.}
    \label{fig:subplots_mcmc}
\end{figure}

\end{minipage}

$ $ \\

$ $ \\

$ $ \\

$ $ \\

$ $ \\

\vspace{20cm}

\begin{minipage}{\textwidth}
\section{Comparing simulations with the observations and model $\mathcal{M}3$} \label{sec:bigbox_fit}

\begin{figure}[H]
    \centering
    \makebox[\textwidth][l]{
        \subfloat[Simulation $\mathrm{const}\lambda$ fitted with model $\mathcal{M}3$]{\includegraphics[width=0.48\textwidth]{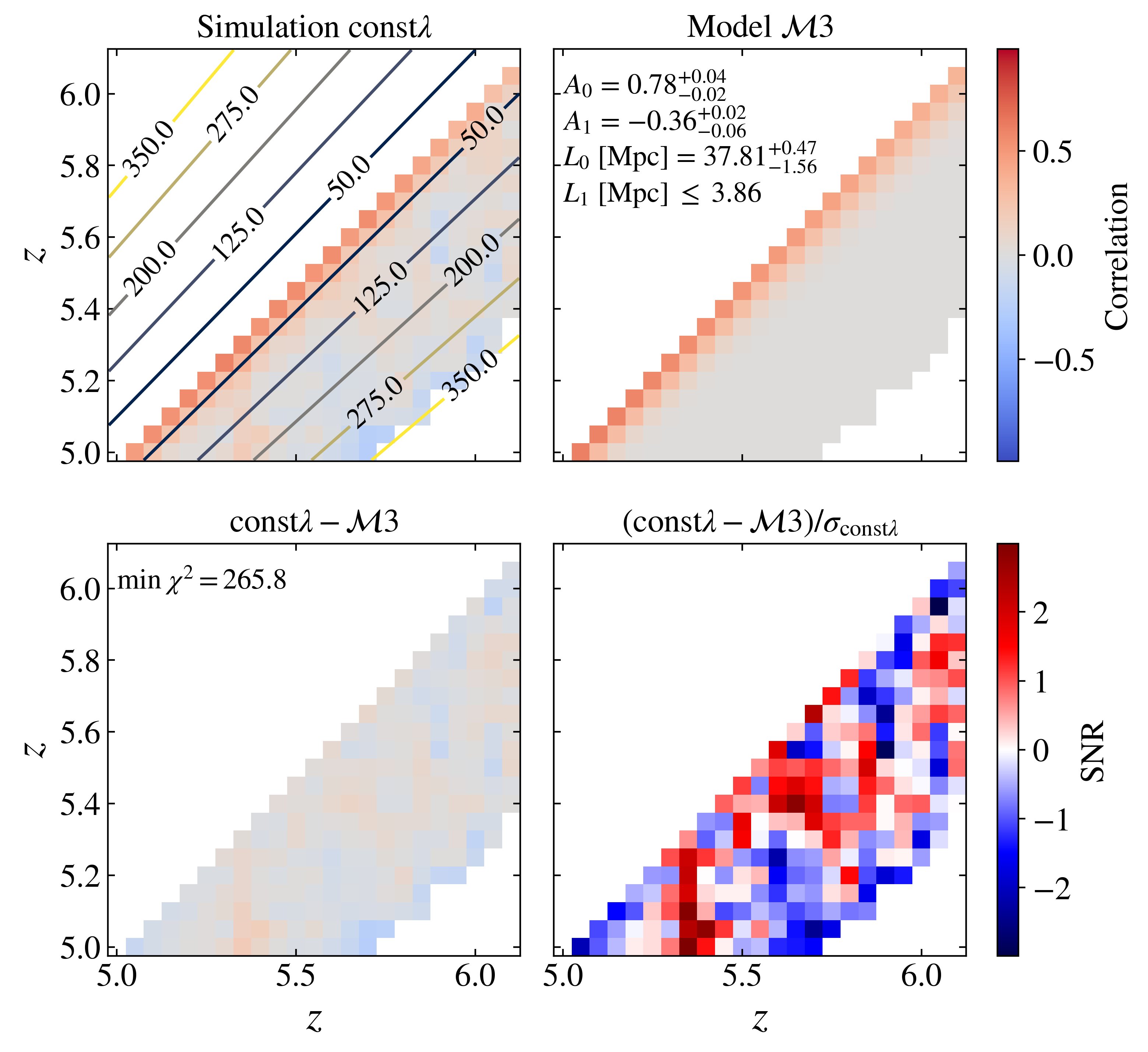}}
        \subfloat[Simulation $\mathrm{evolv}\lambda$ fitted with model $\mathcal{M}3$]{\includegraphics[width=0.48\textwidth]{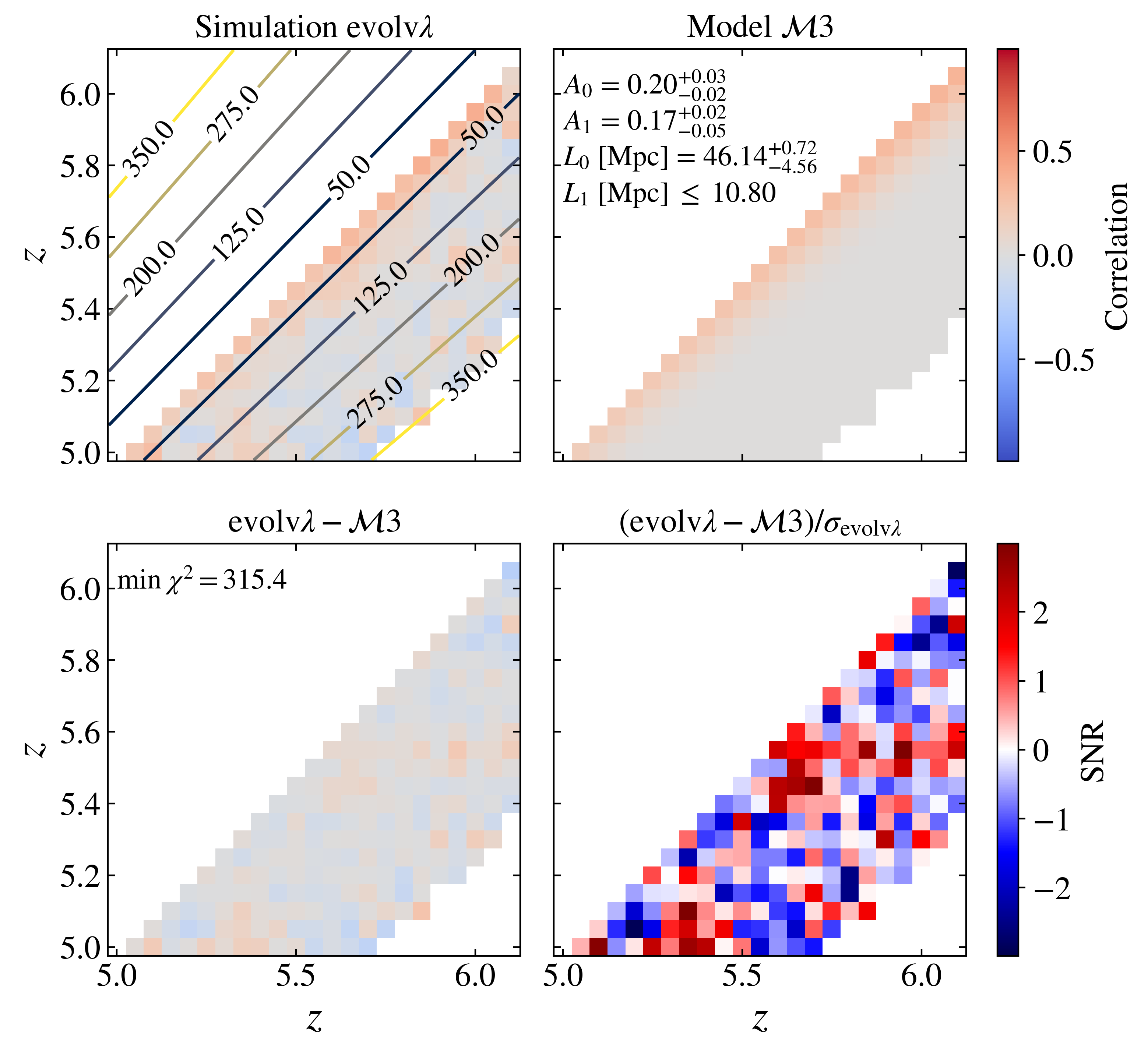}}
    }
    \makebox[\textwidth][c]{
        \subfloat[Simulation $\mathrm{SCRIPT}$ fitted with model $\mathcal{M}3$]{\includegraphics[width=0.48\textwidth]{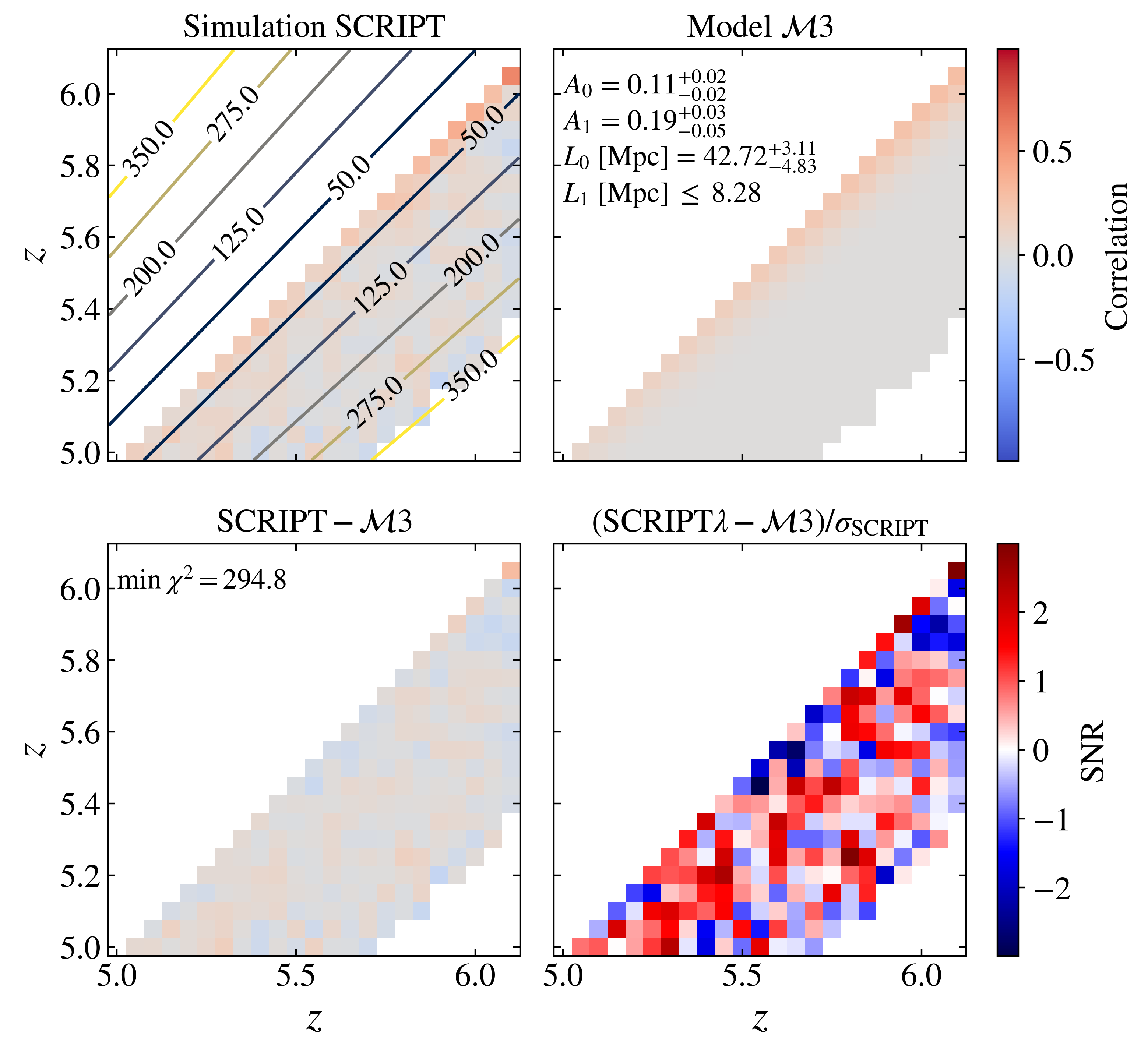}}
    }
    \caption{As Figure~\ref{fig:fit_models_to_data}, but here the different simulations are fitted with model $\mathcal{M}3$. }
    \label{fig:fit_bbbb_with_M3}
\end{figure}

\end{minipage}

\newpage

$ $ \\

$ $ \\

$ $ \\

$ $ \\

$ $ \\

\vspace{20cm}

\begin{minipage}{\textwidth}

\begin{figure}[H]
    \centering
    \makebox[\textwidth][l]{
        \subfloat[Simulation $\mathrm{const}\lambda$ fitted with model $\mathcal{M}3$]{\includegraphics[width=0.48\textwidth]{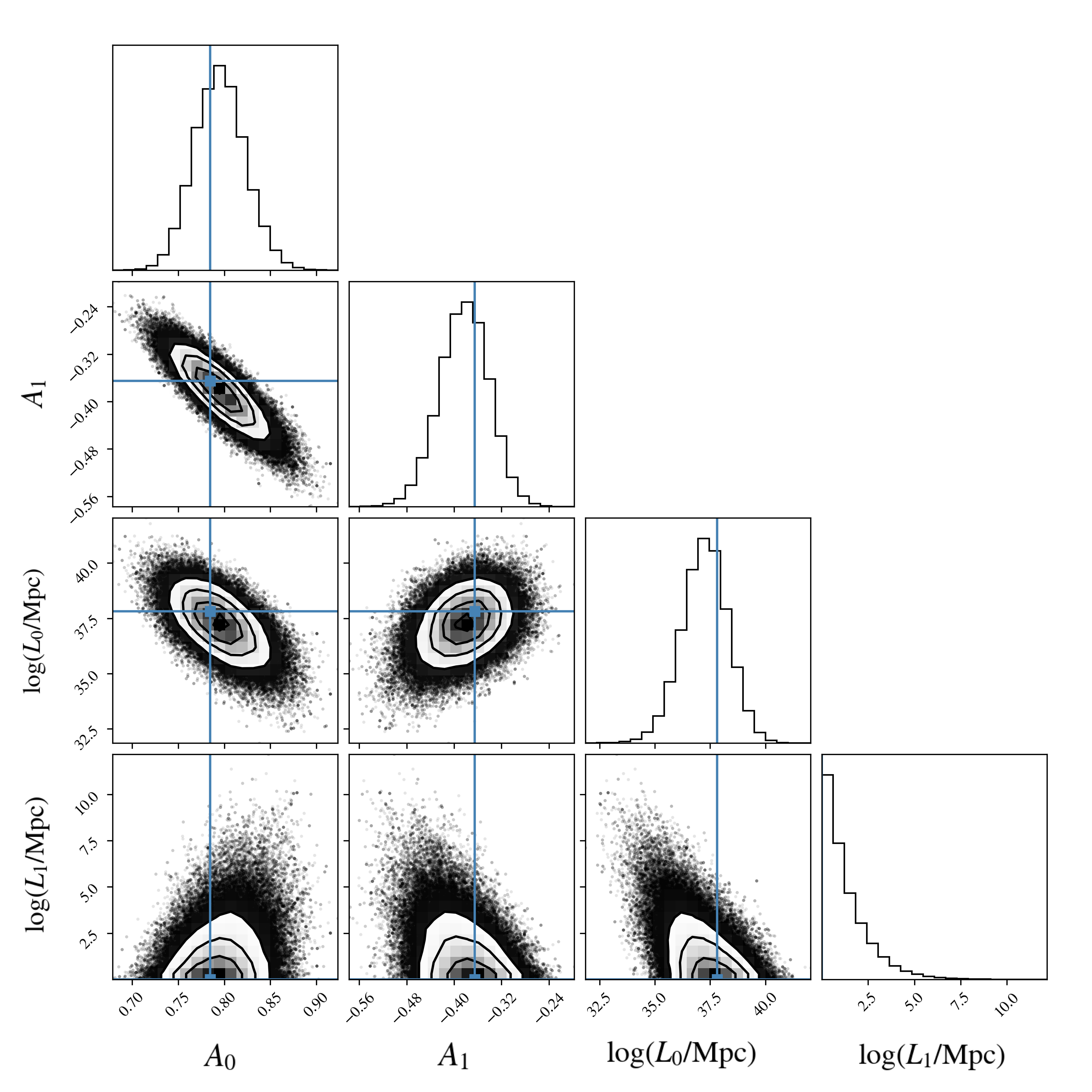}}
        \subfloat[Simulation $\mathrm{evolv}\lambda$ fitted with model $\mathcal{M}3$]{\includegraphics[width=0.48\textwidth]{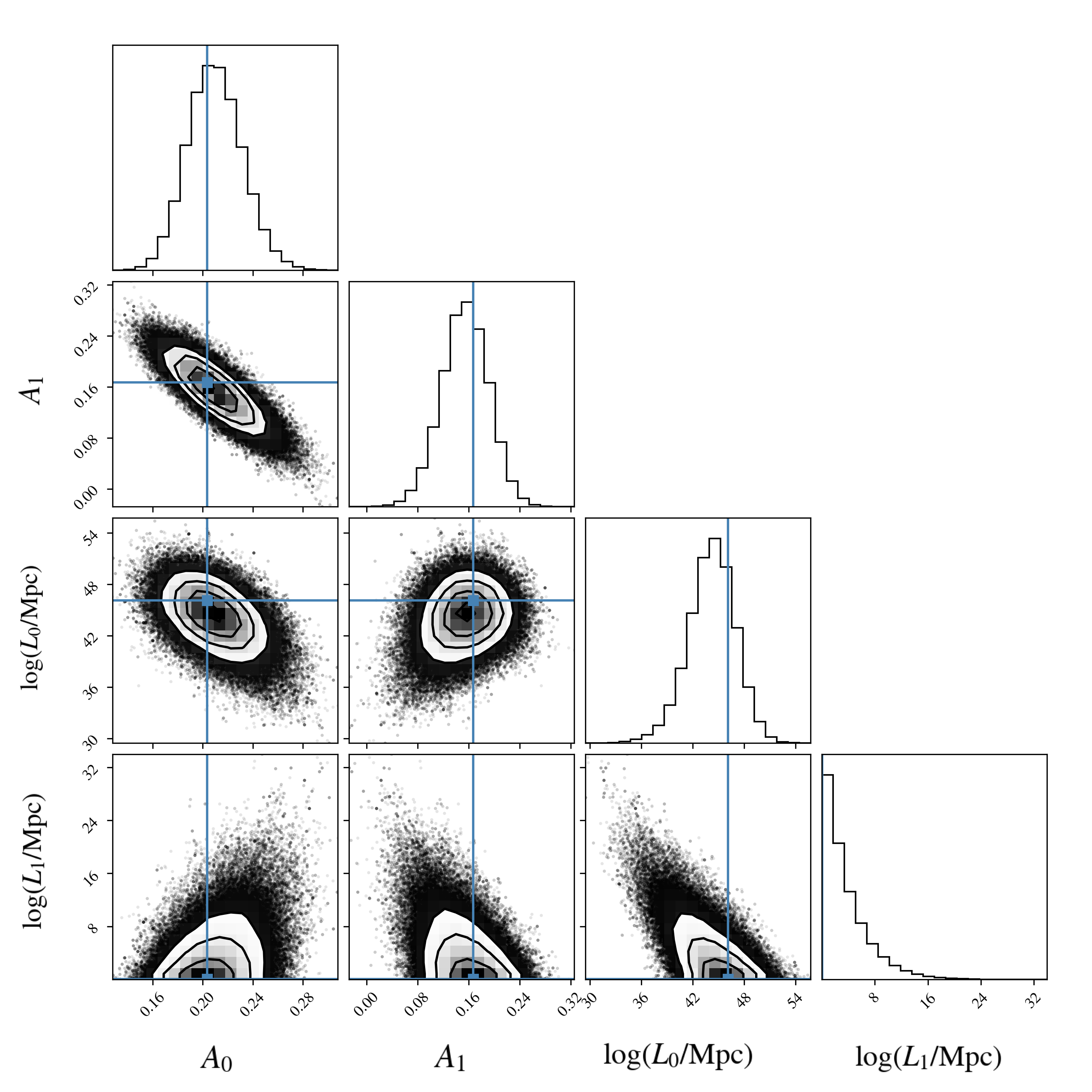}}
    }
    \makebox[\textwidth][c]{
        \subfloat[Simulation $\mathrm{SCRIPT}$ fitted with model $\mathcal{M}3$]{\includegraphics[width=0.48\textwidth]{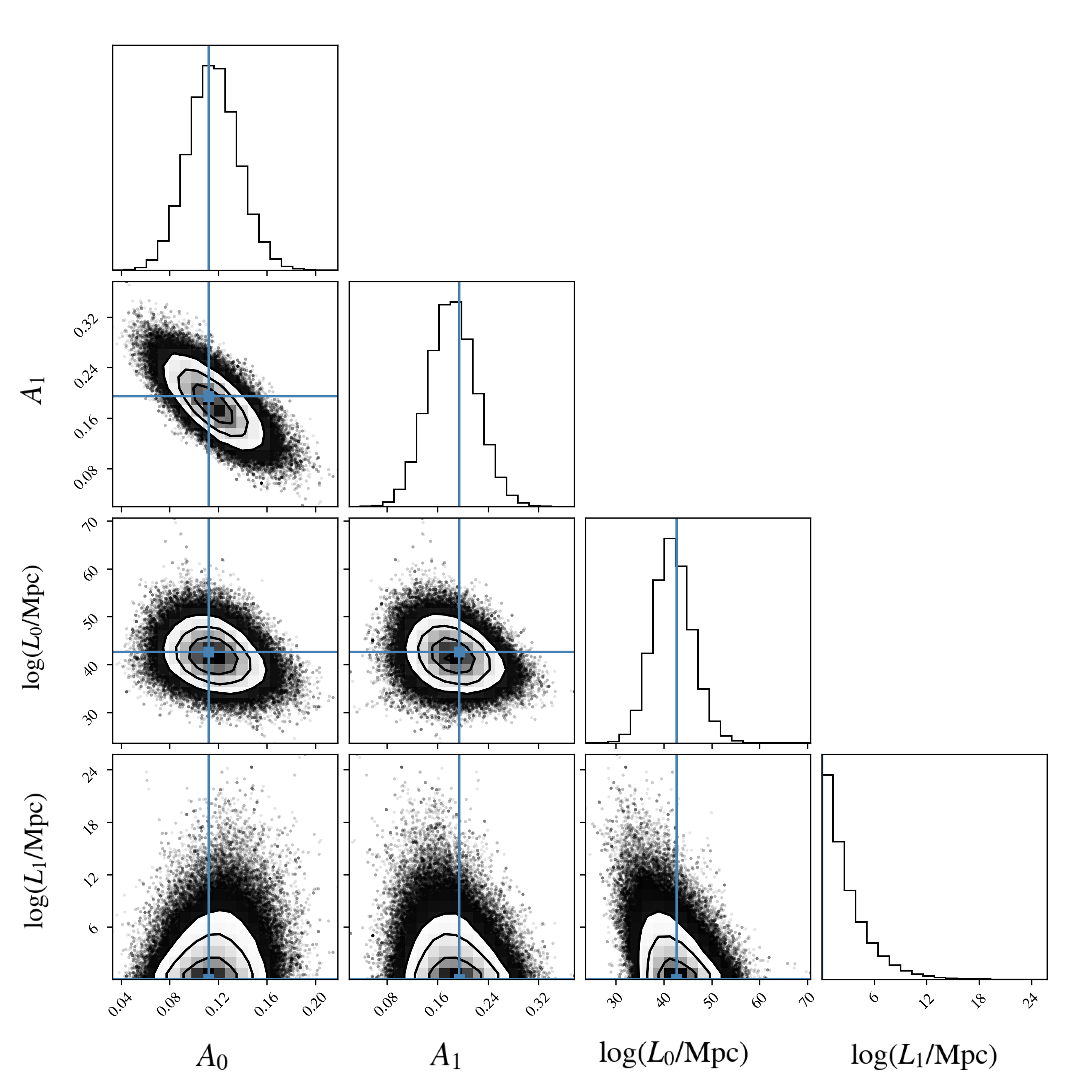}}
    }
    \caption{Posterior distribution for Figure~\ref{fig:fit_bbbb_with_M3}, where the different simulations are fitted with model $\mathcal{M}3$.}
    \label{fig:mcmc_bbbb_with_M3}
\end{figure}

\end{minipage}

\newpage

$ $ \\

$ $ \\

$ $ \\

$ $ \\

$ $ \\

\vspace{20cm}

\begin{minipage}{\textwidth}

\begin{figure}[H]
\centering
\makebox[\textwidth][l]{
\subfloat[BIBORTON: constant $\lambda_\mathrm{mfp}$ \label{fig:fit_S1}]{\includegraphics[width=0.48\textwidth]{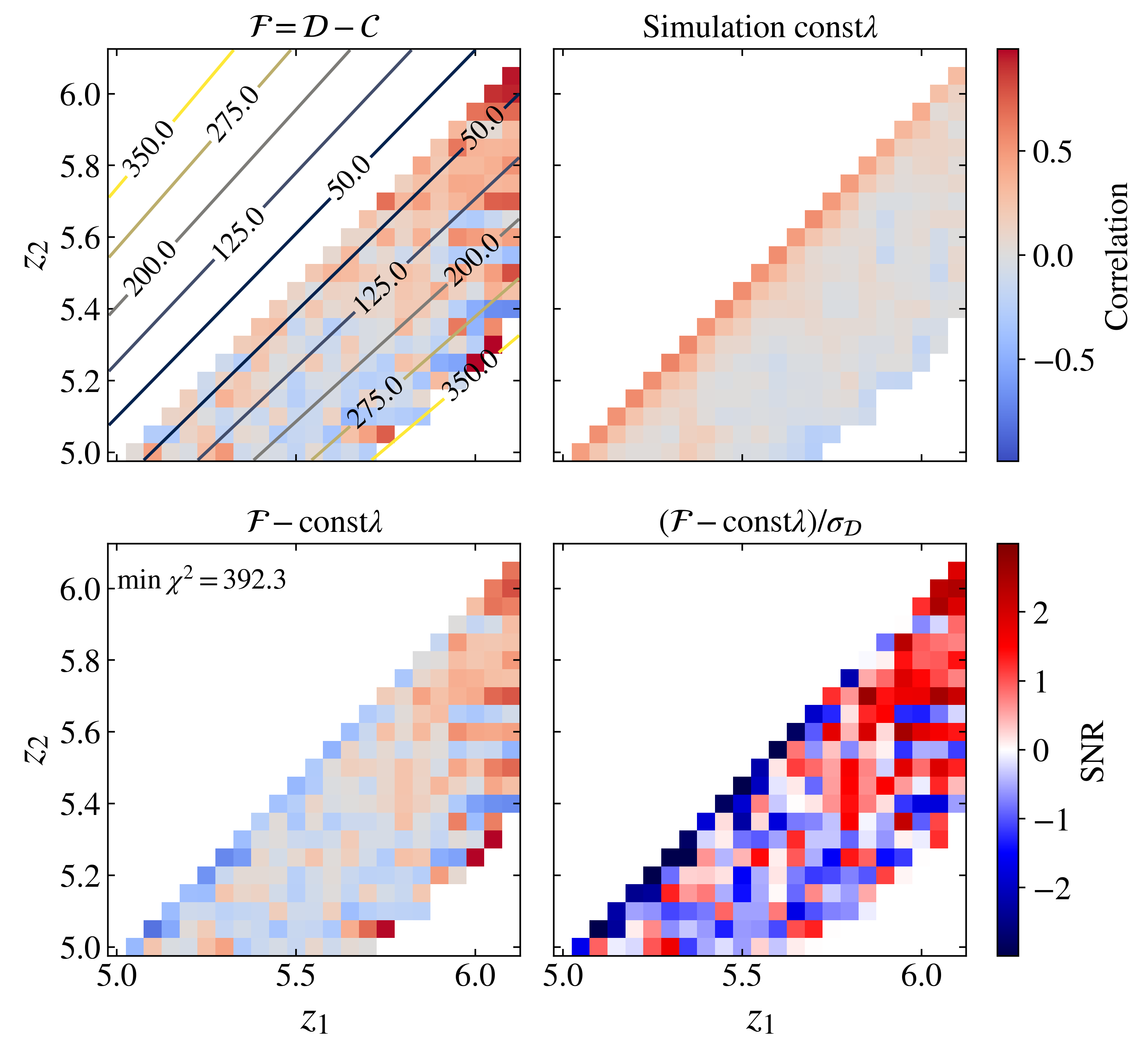}}
\subfloat[BIBORTON: evolving with redshift $\lambda_\mathrm{mfp}$ \label{fig:fit_S2}]{\includegraphics[width=0.48\textwidth]{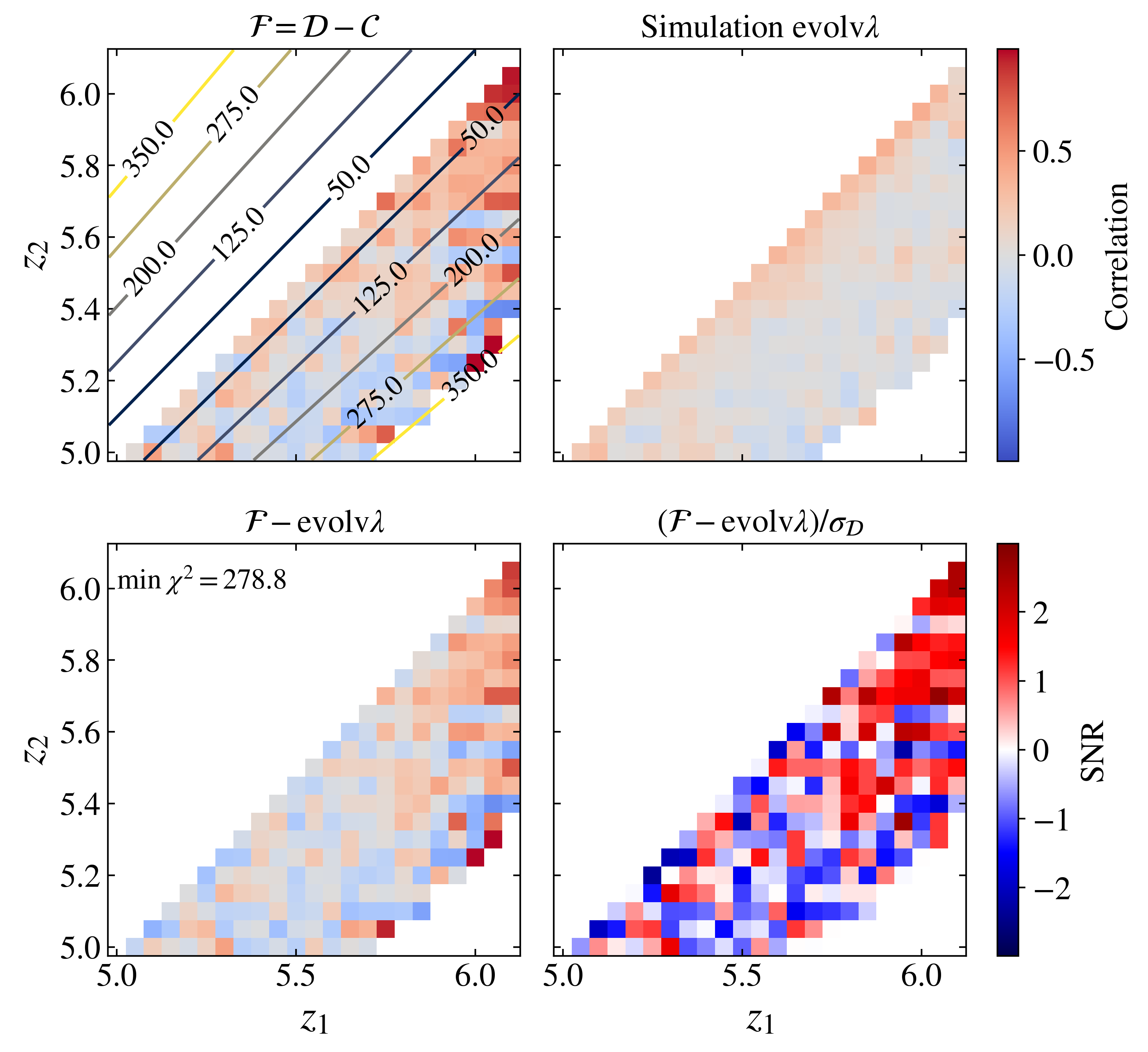}}
}
\caption{As Figure~\ref{fig:fit_models_to_data} for BIBORTON $\mathrm{const}\lambda$ and $\mathrm{evolv}\lambda$ simulations.}
\label{fig:fit_S_appendix}
\end{figure}

\end{minipage}

\newpage

$ $ \\

$ $ \\

$ $ \\

$ $ \\

$ $ \\

\vspace{20cm}

\begin{minipage}{\columnwidth}

\section{Underestimated continuum-reconstruction error} \label{sec:1.5sigma}
\begin{figure}[H]
\makebox[\textwidth][l]{\includegraphics[width=\textwidth]{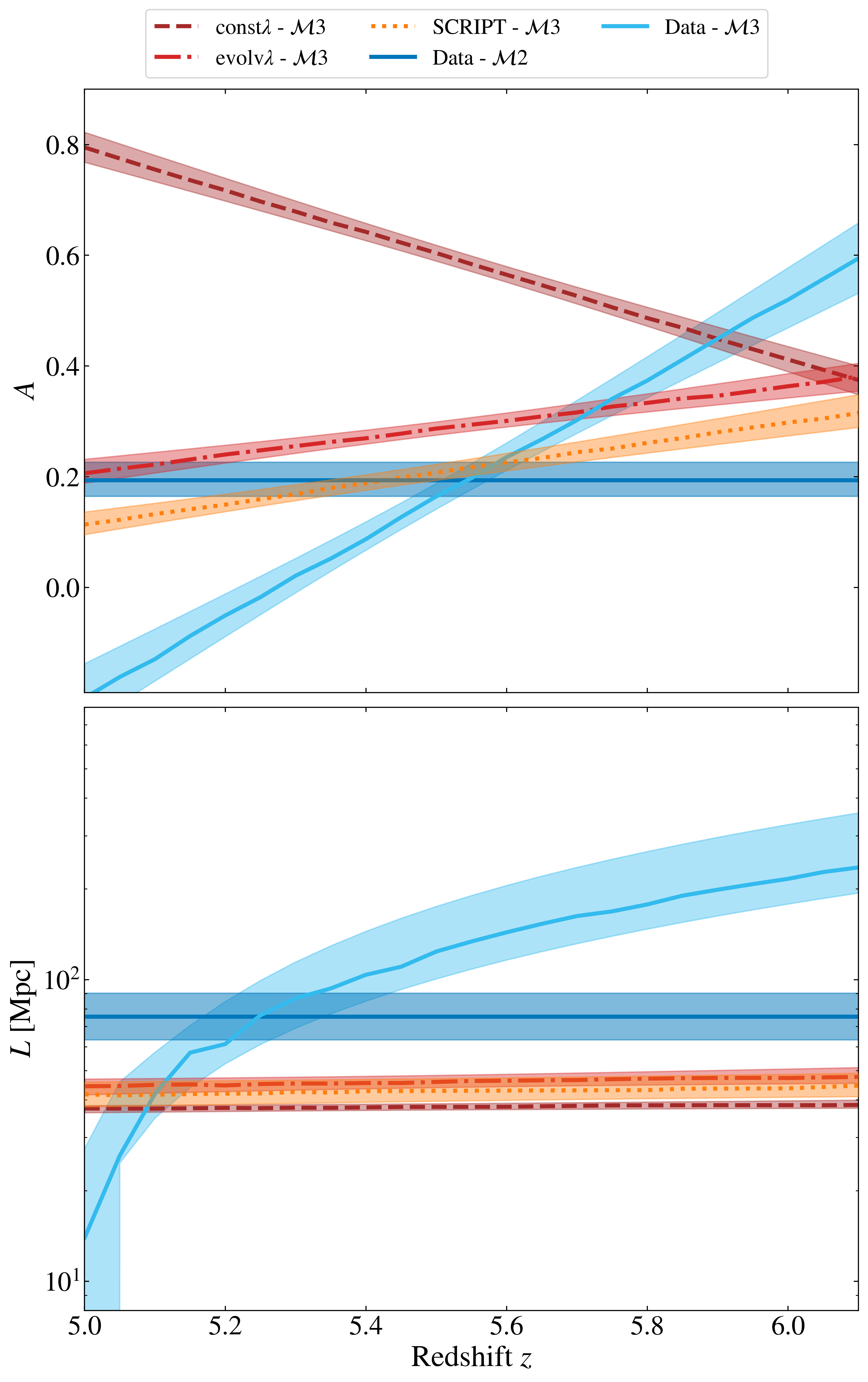}}
\caption{Same as Figure~\ref{fig:fig_tot} but increasing the uncertainty on the continuum-reconstruction by 50\%.}
\label{fig:fig_tot_1.5sigma}
\end{figure}

\end{minipage}

\end{appendix}

\end{document}